\documentclass[sigconf]{acmart}
\AtBeginDocument{%
  }

\copyrightyear{2026}
\acmYear{2026}
\setcopyright{cc}
\setcctype{by}
\acmConference[RecSys '26]{20th ACM Conference on Recommender Systems}{September 27-October 02, 2026}{Minneapolis, MN, USA}
\acmBooktitle{20th ACM Conference on Recommender Systems (RecSys '26), September 27-October 02, 2026, Minneapolis, MN, USA}
\acmDOI{10.1145/3773078.3831824}
\acmISBN{979-8-4007-2284-4/2026/09}

\usepackage{graphicx}
\usepackage{caption}
\usepackage{subcaption}
\usepackage{bm}
\usepackage{tabularx}
\usepackage{multirow}
\usepackage{makecell}
\usepackage{xr}
\usepackage{placeins}
\usepackage{enumitem}
\usepackage{array}
\usepackage{float}
\usepackage{dblfloatfix}
\usepackage{makecell}
\usepackage{amsmath}
\usepackage{algorithm}
\usepackage{algpseudocode}
\usepackage{tikz}
\usepackage{xcolor}

\usepackage{hyperref}       
\usepackage{url}            
\usetikzlibrary{arrows.meta, positioning}

\newcolumntype{C}[1]{>{\centering\arraybackslash}m{#1}}
\fontsize{7.8}{8}\selectfont

\begin{document}
\newcommand{\keywordList}{position bias, unbiased learning to rank,  control function}

\title{A Control Function Framework for Mitigating\\ Position Bias in Learning to Rank Systems}


\author{Md Aminul Islam}
\affiliation{%
  \institution{University of Illinois Chicago}
  \city{Chicago}
  \state{IL}
  \country{USA}
 }
\email{mislam34@uic.edu}

\author{Kathryn Vasilaky}
\affiliation{%
  \institution{California Polytechnic State University}
  \city{San Luis Obispo}
  \state{CA}
  \country{USA}
}
\email{kvasilak@calpoly.edu}

\author{Elena Zheleva}
\affiliation{%
  \institution{University of Illinois Chicago}
  \city{Chicago}
  \state{IL}
  \country{USA}
 }
\email{ezheleva@uic.edu}

\renewcommand{\shortauthors}{Md Aminul Islam, Kathryn Vasilaky, \& Elena Zheleva}

\begin{abstract}

Learning-to-rank (LTR) systems commonly depend on implicit feedback, such as user clicks, because it is easy to collect and can serve as a valuable signal of user preferences. However, directly optimizing ranking models using implicit feedback data often yields suboptimal performance because such data is inherently skewed by systematic biases. Among these biases, position bias is particularly pervasive: items ranked higher tend to receive disproportionately more interactions, regardless of their actual relevance. To address this, we introduce a novel two-stage framework based on control functions. In the first stage, we utilize exogenous variation from the residuals of the ranking process, which are then incorporated into a second stage click model to account for position-dependent distortions. In contrast to existing methods, our approach avoids explicit propensity estimation, supports nonlinear ranking models, and can be flexibly incorporated into any state-of-the-art ranking algorithm for position bias correction. We also propose a debiasing strategy for validation clicks that enables reliable hyperparameter tuning in the absence of unbiased validation data. Empirical results show that our method outperforms state-of-the-art position bias correction methods on both benchmark and real-world industrial datasets.

\end{abstract}


\keywords \keywordList


\maketitle

\section{Introduction} \label{sec:introduction}

Learning-to-rank (LTR) systems have become a core component in many online applications, from search engines to recommender systems. These systems typically use implicit feedback data, such as user clicks, as a proxy for user preferences due to its ease of collection compared to costly manual human annotations. While extremely valuable for learning user preferences at scale, implicit feedback data is prone to biases~\cite{ai-sigir18, yue-www10, joachims-wsdm17}, with position bias being particularly pervasive. Position bias is the tendency of users to interact more with higher-ranked items, regardless of their true relevance~\cite{craswell-wsdm08, joachims-tois07}. This skews click data, causing LTR models to overestimate higher-ranked items and underestimate lower ones. Training on such biased data degrades ranking performance and reinforces a feedback loop that negatively affects user experience.

State-of-the-art methods that address position bias can be broadly categorized as click models, propensity-based approaches, and econometrics-based models. Click models~\cite{chapelle-www09, craswell-wsdm08, dupret-sigir08, wang-www13} estimate true relevance of items by hypothesizing how users browse and maximizing the likelihood of observed clicks.
Click models rely on repeated observations of the same query–item pair to obtain reliable estimates~\cite{mao-sigir19, zou-neurips22}, which limits their effectiveness for tail queries with few interactions.
Two-tower models~\cite{guo-recsys19, yan-sigir22}, following the position-based click model (PBM)~\cite{richardson-www07}, aim to disentangle relevance and bias from clicks by modeling them separately, but may fail when confounding exists between the two~\cite{zhang-kdd23}. Propensity-based methods treat clicks as counterfactual events and re-weight them using inverse propensity weighting (IPW) to achieve relevance-equivalent training objective~\cite{joachims-wsdm17, wang-sigir16}. Propensity estimation often relies on result randomization~\cite{joachims-wsdm17, oosterhuis-sigir20, wang-sigir16}, which can harm user experience~\cite{ai-sigir18, wang-wsdm18}. Some methods~\cite{ai-sigir18, hu-www19, luo-sigir24, vardasbi-cikm20, zhu-recsys20} jointly learn propensities and rankers, but coupling the two raises concerns about the unbiasedness if either model is misspecified~\cite{agarwal-wsdm19, fang-sigir19, tian-sigir20}. 
Doubly robust (DR) IPW-based bias correction methods~\cite{luo-wsdm23, oosterhuis-tis23} have also been proposed recently, where the DR estimator combines an IPW estimate with a regression model and becomes unbiased if either the estimated propensity or the regression model is accurate. Most click modeling and propensity-based methods require modifying the ranker's objective function, which makes them tightly coupled with a specific LTR algorithm and limits their generalization to other LTR algorithms. Alternative approaches~\cite{ovaisi-www20, ovaisi-sigir21}, based on Heckman’s two-stage econometric model~\cite{heckman-econometrica79}, avoid propensity estimation but assume linear ranking models. However, linear models often struggle to capture complex feature interactions and handle the sparse click data typical in LTR systems.

Addressing data bias has a long tradition in econometrics. A prominent class of approaches for addressing various biases is control functions~\cite{wooldridge-jhr15}. This class includes well-known econometric techniques, such as Heckman selection~\cite{heckman-econometrica79} and endogenous switching models~\cite{maddala-75sa}. It also shares the same conceptual foundations as instrumental variables~\cite{reiersol-metrica41} and relies on orthogonality conditions for identification, similar to more recent double machine learning~\cite{chernozhukov-ej18}. Control functions are two-stage methods which model bias through the endogeneity in the error term. In this paper, we introduce a control function method for addressing position bias. In the first stage, we model the system’s previous ranking process, which generated the user feedback data based on observable features. 
In the second stage, we leverage the exogenous variation in first‐stage residuals and feature interaction terms as control functions within the click model to account for position bias. We introduce residual transformation as a hyperparameter in the second stage and suggest several carefully designed transformation options.


A control function approach for addressing position bias offers several advantages over classical Heckman-based selection models~\cite{ovaisi-www20, ovaisi-sigir21}. While Heckman-style methods typically rely on linear outcome equations and strong parametric assumptions, more general control function formulations allow for flexible, potentially nonlinear models in both the selection and outcome stages. Incorporating regularization into the ranking model further mitigates multicollinearity in high-dimensional feature spaces, a challenge that classical Heckman-based implementations are not designed to handle. Unlike propensity-based methods~\cite{ai-sigir18, hu-www19, joachims-wsdm17, luo-sigir24, luo-wsdm23, oosterhuis-sigir20, vardasbi-cikm20, wang-sigir16, wang-wsdm18, zhang-kdd23}, which rely on IPW, our method avoids propensity estimation. Position bias is instead addressed by estimating query-item–specific ranking residuals, without requiring result randomization~\cite{joachims-wsdm17, oosterhuis-sigir20, wang-sigir16}. Moreover, our method corrects for position bias without modifying the ranker’s objective function, enabling application to pointwise, pairwise, and listwise LTR algorithms.

Existing methods are typically evaluated under weak logging policies~\cite{ai-sigir18, hu-www19, ovaisi-sigir21, joachims-wsdm17, luo-wsdm23, oosterhuis-sigir20, ovaisi-www20}, where relevance is typically  assumed to be independent of position. In contrast, real-world systems often operate under strong logging policies~\cite{luo-sigir24}, where position is confounded with relevance~\cite{zhang-kdd23}, and ignoring this mismatch can degrade ranking performance~\cite{luo-sigir24}. Our method remains robust under both weak and strong logging policies and performs well on both benchmark and  industrial datasets. We also address a practical gap in the literature: hyperparameter tuning for an unbiased model when true relevance labels are unavailable. Existing methods~\cite{ai-sigir18, joachims-wsdm17, luo-sigir24, luo-wsdm23, oosterhuis-sigir20, wang-wsdm18} typically assume access to unbiased relevance data or rely on biased clicks for validation, which can lead to suboptimal ranking model~\cite{wang-www24, zhao-www25, niu-sigir25}. We introduce a procedure to debias validation clicks for hyperparameter tuning by removing position-induced effects, ensuring robust performance in the absence of ground-truth relevance.
\vspace{-1.0em}

\vskip -5pt
\section{Related Work}
Position bias in LTR systems is primarily addressed through click models, propensity-based methods, and econometrics-based approaches. Several studies~\cite{chapelle-www09, chuklin-wsdm16, craswell-wsdm08, dupret-sigir08} model clicks via user browsing behavior to infer  
relevance. Craswell et al.~\cite{craswell-wsdm08} propose the Cascade model with sequential examination, while Dupret et al.~\cite{dupret-sigir08} introduce the User Browsing Model (UBM) to estimate position-dependent examination probabilities. Chapelle et al.~\cite{chapelle-www09} propose a Dynamic Bayesian Network (DBN) that captures user satisfaction and search abandonment. Wang et al.~\cite{wang-wsdm18} learn ranking and examination propensity via a graphical model. Guo et al.~\cite{guo-recsys19} propose a two-tower PBM-based model~\cite{richardson-www07} to model relevance and bias.

For propensity-based methods, several approaches~\cite{joachims-wsdm17,oosterhuis-sigir20,wang-sigir16} estimate propensities via result randomization. Wang et al.~\cite{wang-sigir16} use randomized result exposure to estimate query-level position bias and apply IPW, while Joachims et al.~\cite{joachims-wsdm17} show that, under correctly specified propensities, models trained on click data converge to those trained on true relevance labels. Ai et al.~\cite{ai-sigir18} and Hu et al.~\cite{hu-www19} propose jointly learning the propensity and the ranker. Vardasbi et al.~\cite{vardasbi-cikm20} introduce a cascade-model-based method for estimating propensities, and Oosterhuis et al.~\cite{oosterhuis-sigir20} propose a policy-aware propensity approach using stochastic ranking policies to address bias. Zhang et al.~\cite{zhang-kdd23} propose a method to address confounding in item position, but their method is susceptible to under- or over-correcting the confounding effect during propensity estimation which has later been addressed with a two-step optimization model~\cite{luo-sigir24}.
Niu et al.~\cite{niu-cikm25} propose a user-aware IPW estimator that models personalized browsing behavior and aggregates user-specific examination probabilities. Gupta et al.~\cite{gupta-ictir25} propose a two-stage counterfactual learning framework in which a candidate generator retrieves a candidate set of items for a query to reduce the search space, while a re-ranker jointly produces the final ranked list using IPW. Ovaisi et al.~\cite{ovaisi-www20,ovaisi-sigir21} address bias using the Heckman econometric model~\cite{heckman-econometrica79}, avoiding explicit propensity estimation.

Position-biased clicks have also been studied beyond LTR, including IPW-based bandit algorithms for multi-position ad display~\cite{komiyama-icml15, lagree-neurips16}. Several works extend this line using doubly robust (DR) estimators: Saito~\cite{saito-recsys20} for post-click conversion, Kiyohara et al.~\cite{kiyohara-wsdm22} for cascading click-based policy evaluation, Yuan et al.~\cite{yuan-cikm19} for ad click prediction, and Zou et al.~\cite{zou-cikm22} for combining counterfactual learning with DR estimation. However, these methods address problem settings different from unbiased learning to rank.
Recent work by Oosterhuis~\cite{oosterhuis-tis23} and Luo et al.~\cite{luo-wsdm23} propose DR-based bias correction for LTR using IPW. Niu et al.~\cite{niu-sigir25} propose a distributionally robust method, which uses group distributionally robust optimization in existing LTR to handle train–test distribution shifts.


\section{Task Formulation}
We first present the definition of LTR systems following Ai et al.~\cite{ai-sigir18}. Let $\bm{\mathcal{Q}}$ be a universal set of independent and identically distributed (i.i.d.) queries, with each query $q$ sampled from the distribution $P(q)$. Given a query $q$, a ranking model $g \in \mathcal{G}$ predicts a relevance score for each associated item, thereby producing a ranked list for query $q$ by ordering items according to their predicted relevance. When ground-truth relevance annotations are available for all items associated with $q$, we can define $l(g, q)$ as a loss function that quantifies the deviation between the predicted ranking and the true relevance ranking. The goal of an LTR system is to find the optimal model $g$ that minimizes the total empirical loss $\hat{\bm{\mathcal{L}}}(g)$ over all models in $\bm{\mathcal{G}}$.
\begin{equation}
    \hat{\bm{\mathcal{L}}}(g) = \frac{1}{|\bm{\mathcal{Q}}|} \sum_{q \in \bm{\mathcal{Q}}} l(g, q).
\end{equation}
Although human-annotated labels are reliable indicators of true relevance, they are costly to obtain at scale. Instead, LTR systems often rely on implicit feedback, such as user clicks, which is easier to collect and sometimes better reflects user intent~\cite{joachims-kdd02}.
Clicks are binary outcomes, so we model them through the conditional
click probability given query--item features. Let $C_q^y \in \{0,1\}$ denote
whether item $y$ is clicked when shown for query $q$, and let $\bm{x}_q^y$
denote the corresponding query--item features. We can write:
\begin{equation}\label{eq:biased_c}
    \Pr(C_q^y = 1 \mid \bm{x}_q^y) \;=\; g(\bm{x}_q^y),
\end{equation}
where $g(\cdot)$ maps features to click probabilities.
$g(\cdot)$ becomes a \emph{ranking} model when we use
its predicted probabilities (or scores) to order items within a query. Equivalently, we can use a binary response model over regression:
\begin{equation}\label{eq:latent_click}
    C_q^y \;=\; \mathbb{I}\!\left[s(\bm{x}_q^y) + \epsilon^y_{q,g} > 0\right],
\end{equation}
where $s(\cdot)$ is a real-valued score function and $\epsilon^y_{q,g}$ is the error term associated with the model  $g(\cdot)$~\cite{wooldridge2010crosssection}. While we develop the theory with this representation, we can use the same correction terms in other models for clicks, such as standard LTR algorithms.

Although clicks provide feedback on user preferences, they are logged under prior ranking policies. Because users are more likely to click on higher-ranked items, observed clicks reflect both relevance signals in $\bm{x}_q^y$ and item position in the ranking displayed to the user. As a result, naively learning $g(\cdot)$ from observed clicks can
confound feature effects with position effects.

Let $\pi$ be a deterministic ranking policy, and let $r(y \mid \pi_q)$ denote the position of an item $y$ in the system’s previous ranking $\pi_q$ for query $q$. The probability of observing item $y$ depends on its rank~\cite{joachims-wsdm17}, with higher-ranked items receiving more interactions, leading to position bias~\cite{joachims-tois07}. Hence, if position is omitted from the click model equation~\eqref{eq:latent_click}, its influence is absorbed into the
error term $\epsilon^y_{q,g}$.
Since $r(y\mid \pi_q)$ is itself determined by the
historical policy as a function of features, the logged data exhibits a
policy-induced correlation between rank and features. In addition, unobserved determinants of ranking affect exposure and, therefore, clicks, so the error is no longer mean independent of the regressors in the logged click data.
This induces endogeneity in the logged click data, i.e., 
    $\mathbb{E}\!\left[ \epsilon^y_{q,g} \mid \bm{x}_q^y\right] \neq 0$.
This violates the standard exogeneity condition: $\mathbb{E}[\epsilon^y_{q,g} \mid \bm{x}_q^y]=0$,
which is required for consistent estimation of feature effects in binary-response models
(e.g., logit/probit)~\cite{amemiya-1985advanced}.

A naive approach to controlling for position is to include rank as a regressor~\cite{guo-recsys19}, e.g.,  
$C_q^y \;=\; \mathbb{I}\!\left[s\left(\bm{x}_q^y, r(y\mid \pi_q)\right) + \epsilon^y_{q,g} > 0\right]$.
However, because rank is generated by the historical policy using the same features $\bm{x}_q^y$, conditioning on $r(y\mid \pi_q)$
can reintroduce policy-induced variation into the learned feature 
effects. Moreover, the rank or position of an item can be confounded with its underlying relevance (features)~\cite{zhang-kdd23}, especially under strong logging policies~\cite{luo-sigir24}. As a result, rank is not exogenous, and directly conditioning on it leads to biased estimates. Since our goal is to recover feature effects on clicks, which generalizes across ranking policies and out-of-sample queries, unbiased estimation of features requires isolating variation in item position that is \textit{not} explained by feature-driven relevance. This also addresses the confounding between position and relevance induced by the logging policy. Furthermore, rank is not available at prediction time for new queries and items. In the next section, we model the system’s previous ranking process to extract exogenous variation \textit{not} explained by features and use it to account for position bias in the click equation~\eqref{eq:latent_click}.

\section{Position Bias Correction Using Control Function}
We propose Control Function-based Correction (CFC), a two-stage framework for correcting position bias in LTR systems based on a control function approach. Our method uses residuals from the historical ranking process to capture variation in item rankings not explained by observable query–item features. In the first stage, we model the system’s historical ranking process to isolate this feature-unexplained variation. In the second stage, these residuals and their interactions with each feature are incorporated as control function terms in the click model to correct for position bias.



\subsection{Stage 1: Control function for item ranking} \label{control_function_section} 
In the first stage, we model the previous ranking process that generated the training data, where users were more likely to click on
higher-ranked items, leading to position bias in the clicks.
Item relevance depends on observable query–item features, and these features are used by the ranking policy to determine item positions.
Therefore, we model the historical ranking process as:
\begin{equation} \label{eq:control_function}
    r(y \mid \pi_q) = m(\bm{x}_q^y) + \epsilon^{y}_{q,m},
\end{equation}
where $m(\cdot)$ captures the relationship between features and item position under the previous ranking policy. $\epsilon^{y}_{q,m}$ represents unobserved factors that influence ranking outcomes beyond the features (item relevance), such as system biases, limited or noisy training data, randomization, or personalization effects, which influence item placement for a given query but are not explained by observable features. In other words, they capture the component of ranking that is not explained by feature-driven relevance. We refer to this step as \emph{residualization}: decomposing the observed rank into a component explained by the observable features, $m(\bm{x}_q^y)$, and a residual component, ${\epsilon}^{y}_{q,m}$, that captures feature-unexplained variation in the ranking process. An important advantage of this 
control function model is that $m(\cdot)$ can be estimated using any linear or nonlinear model~\cite{blundell-res04, das-res03}. This flexibility allows the first-stage model to accommodate high-dimensional feature interactions and to incorporate regularization, resulting in more stable estimates in settings where classical parametric selection models may fail. 

After fitting the ranking equation~\eqref{eq:control_function} based on available data where each item was placed in the historical ranking, we estimate residuals for each query-item pair. Let $\hat{\epsilon}^{y}_{q,m} = r(y \mid \pi_q) - \hat{m}(\bm{x}_q^y)$ 
denote the estimated ranking residual. By construction, when $m(\cdot)$ captures all systematic feature-driven variation in rank, the resulting residuals are mean-independent of the features, i.e., 
$\mathbb{E}[\epsilon^{y}_{q,m} \mid \bm{x}_q^y] = 0$. In the second
stage, we will use the estimated residuals to control for variation in the ranking process—beyond what is explained by observable features. 
This follows the control function logic to absorb dependence between the ranking process and the click outcome that arises from the historical ranking policy.

It is important to note that residualization is necessary but not sufficient to recover unbiased feature parameters. Partialing the features out of rank removes the component of position that is predictable from $\bm{x}_q^y$. However, because $\hat{\epsilon}^{y}_{q,m}$ is itself computed from the observed rank and those same features, it introduces no additional variation in position that is independent of the factors driving user clicks. Without variation of this kind, position bias cannot be disentangled from feature-driven relevance.




This is analogous to the residualization step used in the first stage of double machine learning~\cite{chernozhukov-ej18}, where predictable components of the treatment variable are partialed out with respect to observed covariates. In that setting, partialing out is sufficient because the observed covariates are assumed to account for all confounding affecting the treatment variable. That assumption does not hold in our setting, since position bias is already embedded in the features used for ranking. Consequently, residualization alone is insufficient and should be paired with an additional source of independent variation.


In our setting, the identifying variation comes from feature-dependent heteroskedasticity in the ranking errors, that is, ranking uncertainty that varies systematically with the features. For example, rankings may be noisier for infrequent queries, newly introduced items, or items with limited interaction history. Following the identification theory of Lewbel~\cite{lewbel-jbec12}, we exploit this heteroskedasticity to construct the additional control function terms defined in Section~\ref{subsec:stage2}. These terms provided the feature-dependent variation in the ranking process used in Stage 2. Formally, we say that the ranking errors are feature-dependent heteroskedastic if 
$\operatorname{Var}(\epsilon^{y}_{q,m} \mid \bm{x}_q^y)$ varies systematically with the features, namely, \
$\operatorname{Var}(\boldsymbol{\epsilon}_m \mid \mathbf{X})
= \operatorname{diag}\!\left(\sigma_m^2(\bm{x}_{q_1}^{y_1}), \ldots, \sigma_m^2(\bm{x}_{q_n}^{y_n})\right)$. 
We show the presence of heteroskedasticity in the ranking residuals across common LTR benchmark datasets (see Subsection~\ref{sec:experimental_setup}).

\subsection{Stage 2: Click model correction} \label{subsec:stage2}


In the second stage, we train the click model while conditioning on residual variation from the first-stage ranking process. Following control function theory~\cite{wooldridge-jhr15}, endogeneity arises because unobserved factors affecting the historical ranking policy are correlated with unobserved determinants of clicks. Conditioning on a function of the ranking residual controls for this dependence.

The simplest such specification conditions on the ranking residual alone, e.g.,
$C_q^y \;=\; \mathbb{I}\!\left[s(\bm{x}_q^y, \hat{\epsilon}^{y}_{q,m}) + \epsilon^y_{q,g} > 0\right]$.
As discussed in Section~\ref{control_function_section}, this residual-only specification cannot recover unbiased feature parameters on its own without additional distributional assumptions. Nevertheless, we include it as an empirical baseline (CFC-S) in our experiments.
We instead follow Lewbel~\cite{lewbel-jbec12} by augmenting the control function with additional terms constructed from feature-dependent heteroskedasticity in the rank equation. This approach is well motivated in LTR settings where the variance of ranking errors naturally depends on the query and item features.


Following Lewbel, the resulting click model can be written as:
\begin{equation}\label{eq:unbiased_c_lewbel_temp}
    C_q^y \;=\; \mathbb{I}\!\left[s\!\left(\bm{x}_q^y,\ \Phi(\bm{x}_q^y,\hat{\epsilon}_{q,m}^y)\right) + \epsilon^y_{q,g} > 0\right],
\end{equation}
where $\Phi( \bm{x}_q^y, \hat{\epsilon}_{q,m}^y)
= (\bm{x}_q^y - \bar{\bm{x}}_q^y) \odot \hat{\epsilon}_{q,m}^y$ are the
control function terms, with \( {\bar{\bm{x}}_q^y} = (\bar{x}_1, ..., \bar{x}_d) \) denoting the mean of each feature dimension, and \( \odot \) representing element-wise multiplication. Therefore, each feature dimension is mean-centered within the query and then interacted
with the ranking residual. Intuitively, these interactions capture feature-dependent variation in the ranking noise that shifts item positions under the historical policy, and conditioning on them helps absorb remaining policy-induced dependence in the logged click data without conditioning directly on rank.

\textbf{Identification in our control function framework.} The recovery of unbiased parameters using Lewbel's method is valid under the following required conditions~\cite{wooldridge2010crosssection, lewbel-jbec12}:

\begin{itemize} [leftmargin=10pt, nosep]
    \item \textit{Heteroskedasticity of the ranking residuals:} The ranking errors should be feature-dependent heteroskedastic. As mentioned earlier, we empirically demonstrate the presence of heteroskedasticity in the residuals across common LTR benchmark datasets.
    
    \item \textit{Standard covariate exogeneity:} The observed covariates (features) should be exogenous, meaning the covariates are uncorrelated with the structural error in the click equation, i.e.,
        $\mathbb{E}[\bm{x}_q^y \epsilon^{y}_{q,g}] = 0$.
    In historical click data, the observed query-item features are determined prior to exposure, user interaction, and ranking. As a result, the exogeneity assumption is appropriate in our setting since the features are fixed before any ranking or clicking decisions occur, and are therefore plausibly uncorrelated with the structural error in the click equation. This assumption requires only that the features reflect information available before the production ranker generates the ranking in the historical click data. Consequently, even features such as historical click-through rate, popularity, or prior exposure satisfy the assumption, provided they are computed using information available before the historical ranking is produced.
    
    \item \textit{Exogeneity with respect to error interaction terms:} The features should be uncorrelated with any joint dependence between the unobserved variation in the ranking and click equations, i.e., 
        $\operatorname{Cov}\!\left(\bm{x}_q^y, \epsilon^{y}_{q,m}\epsilon^{y}_{q,g}\right) = 0.$
    This assumption is applicable in LTR settings because features are determined prior to exposure and cannot depend on unobserved shocks that jointly affect ranking and clicking behavior.
\end{itemize}
Under the latter two assumptions, heteroskedasticity provides the variation needed to account for remaining rank-related correlation, while feature exogeneity ensures that the correction does not introduce additional bias. Because all these assumptions are satisfied in our LTR settings, CFC is theoretically grounded to recover unbiased feature parameters under Lewbel’s identification framework.

CFC can operate under both deterministic and stochastic ranking policies. In deterministic settings, where item positions are correlated with feature-driven relevance, first-stage residuals capture variation in rankings not explained by observed features and act as control functions in the second stage to correct for position-dependent endogeneity induced by the historical ranking policy. In stochastic logging policies, randomized exposure introduces variations in item positions, leading to more  variations in ranking residuals, which further facilitates the
consistent  estimation of feature parameters. We empirically demonstrate the effectiveness of our method in both deterministic and stochastic logging policies.

We train the second-stage model using a learning-to-rank (LTR) algorithm, taking the features and control function terms as inputs and item clicks as targets. A key advantage of our framework is its flexibility: the second-stage model can be instantiated with any linear or nonlinear, pointwise, pairwise, or listwise LTR algorithm. Although we formulate the second stage as a binary response model because click data is binary, control function methods are not restricted to binary outcomes. They naturally extend to continuous-response settings and accommodate nonparametric first-stage learners. Furthermore, the binary response formulation is fully compatible with modern LTR algorithms, which can be trained using either binary click data or graded relevance labels.

After learning the optimal model, we can predict click probabilities for new queries using the item features. The control function terms are required during training to break the dependence between item position and the click error, ensuring consistent estimation of feature effects on clicks. Once these unbiased feature parameters are identified, prediction for new queries uses the learned model, relying on the relevance component (i.e., estimated feature parameters) to obtain unbiased click probabilities. Hence, at inference time for new queries, the control function terms are simply set to zero, i.e., $\Phi(\cdot)=\mathbf{0}$. Formally, for a new query $q$ and item $y$, this corresponds to: $\mathbb{E}[C^{y}_{q} \mid \bm{x}_q^y] = \Pr(C_q^y = 1 \mid \bm{x}_q^y) = g(\bm{x}_q^y)$, where $g(\cdot)$ maps features to click probabilities. Thus, at inference time, we ask what score the model would assign to an item for a query if there were no residual shock.

\textbf{Contrast with other methods for position bias correction}.
Propensity-based models require estimating examination probabilities~\cite{ai-sigir18} using result randomization~\cite{joachims-wsdm17, oosterhuis-sigir20, wang-sigir16} or offline modeling~\cite{luo-wsdm23, luo-sigir24}, and then use these propensities to reweight clicks  to get an unbiased ranking loss~\cite{ai-sigir18}. On the other hand, basic click models ~\cite{chapelle-www09, craswell-wsdm08, mao-sigir18} typically aim to separate the effects of item position (bias) and item relevance (feature effects), and use the relevance component for an unbiased ranking model. Therefore, unlike propensity-based methods that rely on explicitly modeling the exposure mechanism or click models that disentangle the relevance component, CFC uses variations in first-stage ranking residuals to account for position-dependent distortions in the second stage.

While our proposed method and Heckman-based position bias correction belong to the broader class of control function techniques, methods in this family can differ substantially in their assumptions and applications, reflecting the diversity of econometric approaches encompassed by the control function framework. The assumptions needed for unbiased estimation of feature parameters in our method are less stringent than for the specific sub-case of Heckman-based correction methods~\cite{ovaisi-www20, ovaisi-sigir21}. The Heckman model relies on linearity assumptions in both stages and joint distributional 
assumptions for unbiased estimation of feature parameters. Linear models are unsuitable for LTR tasks that involve high-dimensional, non-linear feature interactions, and highly sparse clicks. Using modern nonlinear rankers violate the assumptions of Heckman methods. On the other hand, our control function approach, which leverages the feature-dependent heteroskedastic ranking residuals from the previous ranking process, does not require linearity assumptions in either stage and can handle multicollinearity in high-dimensional feature spaces by selecting flexible models with regularization. More recent double machine learning~\cite{chernozhukov-ej18} differs from the control function approach because it estimates an unbiased causal effect through residual-on-residual orthogonalization, whereas our method leverages feature-dependent heteroskedastic ranking residuals.

\subsection{Modeling the residuals}
There are several ways to model the first-stage residuals to construct control function terms in the second-stage. Most studies directly incorporate residuals in the second stage 
but a few also show the value of using residual transformations, such as 
converting the first stage residuals into a likelihood score~\cite{rivers-econometrica88} and estimating a kernel function of the residuals~\cite{blundell-res04}.
Instead of committing to a single residual transformation, we treat the transformation itself as a hyperparameter in training the second-stage model—an option that, to our knowledge, has not been explored in prior work. This allows us to optimize the transformation for ranking performance using validation data. We propose four carefully designed transformations of the residuals, each capturing a unique aspect of position bias correction, but our framework can easily incorporate others:
\begin{enumerate}[leftmargin=12pt, nosep]
    \item \textbf{Min-Max normalization:} We apply min-max normalization to the residuals ($\hat{\epsilon}^{y}_{q,m}$), and use the normalized values in the model (equation~\eqref{eq:unbiased_c_lewbel_temp}) instead of the raw residuals. This aligns with typical feature normalization and avoids disproportionate impact from the residuals.

    \item \label{PDF-Normal} \textbf{PDF transformation (assuming normal 
    distribution):} We estimate the probability density function (PDF) of the residual ($\hat{\epsilon}^{y}_{q,m}$) for each query-item pair and use the PDF in the click model (equation~\eqref{eq:unbiased_c_lewbel_temp}) instead of using the raw residuals directly. The PDF captures how typical or unusual a residual is for a query-item pair relative to the overall residual distribution. In contrast, residuals are absolute measures and do not capture the distributional information of unobserved factors. Therefore, instead of using raw residuals, the PDF can be used to improve the consistency of the ranker’s estimation by capturing distributional information not reflected in the raw residuals.

    \item \label{IMR-normal} \textbf{Hazard ratio 
    transformation (assuming normal distribution):} This transformation under the normality assumption is also known as inverse Mills ratio (IMR). We transform the residual ($\hat{\epsilon}^{y}_{q,m}$) for each query-item pair into a hazard ratio. The hazard ratio is the ratio of the PDF to the cumulative distribution function (CDF). Let $\psi^{y}_{q}$ be the hazard ratio of the residual $\hat{\epsilon}^{y}_{q, m}$ for item $y$ with query $q$, then the hazard function is defined as:
        $\psi^{y}_{q} = \frac{f(\hat{\epsilon}^{y}_{q, m})}  {F(\hat{\epsilon}^{y}_{q, m})}$,
    where $f(\cdot)$ and $F(\cdot)$ are PDF and CDF. We use the 
    estimated IMR in the click model (equation~\eqref{eq:unbiased_c_lewbel_temp}) instead of using the residuals directly. We choose IMR because it is used in regression models to address non-random missing data. Items at lower ranks are less likely to be observed and clicked, even when relevant, leading to missing relevance feedback. This missingness is not random as it depends on the position of the item, which, in turn, is influenced by previous ranking process. Hence, the observed clicks form a 
    non-random, biased sample, introducing position bias in the click data. The IMR accounts for the non-random missingness of items by modeling the probability that an item is selected into a sample based on the underlying distribution of residuals. IMR has also been used in Heckman-based methods~\cite{ovaisi-www20, ovaisi-sigir21} to model the item selection probability for correcting selection bias, while we apply the IMR of the estimated residuals from item ranking process to correct for position bias. 
    The difference between the PDF transformation and this transformation is that the latter is scaled by the CDF.
    
    \item \textbf{Hazard ratio transformation:} Instead of assuming normality as in the PDF and IMR transformations, we use a more general nonparametric Kernel Density Estimation (KDE) to estimate the PDF ($f(\hat{\epsilon}^{y}_{q, m})$) of each residual ($\hat{\epsilon}^{y}_{q,m}$).
    The estimated PDF of the residual is used to calculate the hazard 
    ratio by dividing it by the CDF. We then use the estimated hazard ratio in the model (equation~\eqref{eq:unbiased_c_lewbel_temp}) instead of using the residuals directly. Note that, nonparametric control function 
    methods~\cite{blundell-res04, das-res03} do not require parametric assumptions about the distribution of residuals. 
\end{enumerate}
Let $T(\hat{\epsilon}^{y}_{q,m}) \in \mathcal{T}$ be any transformation of the estimated residuals, with $\mathcal{T}$ being the set of all transformations. Now, we can rewrite the click model from equation~\eqref{eq:unbiased_c_lewbel_temp} as:
\begin{equation}\label{eq:unbiased_c_lewbel}
    C_q^y \;=\; \mathbb{I}\!\left[s\!\left(\bm{x}_q^y,\ \Phi\left( \bm{x}_q^y, T(\hat{\epsilon}_{q,m}^y) \right) \right) + \epsilon^y_{q,g} > 0\right].
\end{equation}
We provide the full CFC algorithm in the Appendix (Algorithm~\ref{alg:cfc_bias_correction}).

\subsection{Debiasing clicks for hyperparameter tuning} \label{sec:debiased_clicks}
In typical machine learning, train, validation, and test data are assumed to come from the same distribution (i.i.d. assumption)~\cite{wang-www24, zhao-www25}. In unbiased learning to rank, however, training uses click data, while proper evaluation relies on unbiased test data. To tune hyperparameters for an unbiased ranking model, it is ideal to use unbiased ground truth for validation. However, such data is rarely available in real-world scenarios and when the validation set is sampled from the training data, as is typically the case, hyperparameter tuning relies on biased click data, which is suboptimal. 
Many existing works on unbiased learning to rank~\cite{ai-sigir18, joachims-wsdm17, luo-sigir24, luo-wsdm23, oosterhuis-sigir20, wang-wsdm18} overlook the issue of appropriate validation in the absence of unbiased data, often either assuming access to unbiased relevance judgments or relying on biased clicks for validation.
To address the problem of hyperparameter tuning when true item relevance is unavailable for validation, which is likely to occur in real-world LTR applications, we propose a method for debiasing click data.

We model the click data as dependent on the features $\bm{x}_q^y$ and the corresponding estimated transformed residual $T(\hat{\epsilon}^{y}_{q,m})$. 
$T(\hat{\epsilon}^{y}_{q,m})$ accounts for the dependence between item position and click error in the click model to account for item position.
To debias the clicks from the position effect, we can partial out the effect of $T(\hat{\epsilon}^{y}_{q,m})$ from clicks. To achieve this, we consider $T(\hat{\epsilon}^{y}_{q,m})$ as the only independent variable affecting clicks, and learn a function $D(\cdot)$ as follows:
\begin{equation} \label{eq:valid_clicks}
    C^y_q = D\left(T(\hat{\epsilon}^{y}_{q,m})\right) + \epsilon^{y}_{q,D},
\end{equation}
where $\epsilon^{y}_{q,D}$ is the error term for item $y$ associated with query $q$ for the function $D(\cdot)$. Therefore, the effect of the features on a click is absorbed in the error term $\epsilon^{y}_{q,D}$. Using the predicted clicks from the function $D(\cdot)$, we calculate the estimated residual $\hat{\epsilon}^{y}_{q,D}= C^{y}_{q} - \hat{C}^{y}_{q}$ for each item $y$ associated with query $q$ in the validation dataset, 
where $\hat{C}^{y}_{q}$ is the predicted value from $D(\cdot)$ for item $y$ associated with query $q$. The estimated residual $\hat{\epsilon}^{y}_{q,D}$ represents the parts of a click that are  separated from the position effect, capturing only unbiased features' effect on a click. Thus, when validation data with true relevance is not available, we can use $\hat{\epsilon}^{y}_{q,D}$ as a proxy for true relevance labels during validation. Our debiasing procedure corrects only for position bias, leaving other sources of bias or noise in the click data outside the scope of this work. 

\section{Experiments}
We aim to address the following research questions in evaluation:
\begin{itemize}[leftmargin=9pt, nosep]
    \item \textbf{RQ1:} How does CFC perform compared to 
    existing position bias correction methods in LTR?
    \item \textbf{RQ2:} Is CFC robust to different degrees of position bias severity?
    \item \textbf{RQ3:} How does CFC perform on real-world click log data?
    \item \textbf{RQ4:} How does CFC perform when the true relevance of the validation data is unavailable?
    \item \textbf{RQ5:} How does the performance of CFC change as the number of clicks increases?
    \item \textbf{RQ6:} Is CFC robust to different levels of click noise?
    \item \textbf{RQ7:} How does CFC perform under stochastic logging policy induced by randomized interventions?
\end{itemize}

\noindent The code for the experiments is available at \url{https://github.com/edgeslab/CFC.git}.

\begin{table*} [b]
  \setlength{\tabcolsep}{4.8pt} 
  \fontsize{7.8}{7}\selectfont
  \captionsetup{justification=raggedright, width=1.0\textwidth}
    \caption{A comparison of the overall performance of CFC and other baselines. CFC and PIJD can be implemented with both LambdaMART and DNN. PairD is coupled with LambdaMART, while UPE, MULTR, PAL, REM, and DLA are implemented with DNN rankers and do not have LambdaMART implementations. The best-performing method for each ranker, excluding Oracle, is shown in bold, and the second-best is underlined. * denotes a statistically significant improvement of the best-performing method over the best baseline, considering all LambdaMART and DNN rankers, with ${\bm{p < 0.05}}$.}
  \vspace{-0.05em}
  \centering
    \begin{tabular}{cccccccccccccccc}
    \toprule
    \multirow{2}{*}{Datasets} & \multirow{2}{*}{Metrics} & \multicolumn{5}{c} {LambdaMART} & \multicolumn{8}{c}{DNN} \\ \cmidrule(r){3-7} \cmidrule(r){8-16}
        & & CFC & PIJD & PairD & LambdaMART & Oracle & CFC & PIJD & UPE & MULTR & PAL & REM & DLA & DNN & Oracle \\
     \midrule

    \multirow{2}[0]{*}{Yahoo}
    & ERR@10 & \textbf{0.462}${^*}$ & \underline{0.455} & 0.452 & 0.447 & 0.474 & \textbf{0.456} & 0.448 & 0.449 & \underline{0.453} & 0.446 & 0.452 & 0.452 & 0.444 & 0.464\\
    & NDCG@10 & \textbf{0.801}${^*}$ & 0.788 & \underline{0.789} & 0.786 & 0.817 & \textbf{0.794} & 0.787 & 0.783 & \underline{0.789} & 0.786 & 0.775 & 0.788 & 0.782 & 0.805\\
    \midrule

    \multirow{2}[0]{*}{\makecell{MSLR-\\WEB10k}}
    & ERR@10 & \textbf{0.324}${^*}$ & 0.277 & \underline{0.285} & 0.270 & 0.342 & \textbf{0.306} & 0.260 & 0.265 & \underline{0.303} & 0.252 & 0.277 & 0.301 & 0.256 & 0.325\\
    & NDCG@10 & \textbf{0.489}${^*}$ & 0.440 & \underline{0.464} & 0.457 & 0.498 & \textbf{0.467} & 0.451 & 0.422 & \underline{0.462} & 0.440 & 0.426 & 0.457 & 0.448 & 0.476\\
    \midrule

    \multirow{2}[0]{*}{Istella-S}
    & ERR@10 & \textbf{0.717}${^*}$ & 0.681 & 0.690 & \underline{0.694} & 0.737 & \textbf{0.689} & 0.674 & 0.645 & \textbf{0.689} & 0.675 & 0.679 & 0.681 & 0.672 & 0.702\\
    & NDCG@10 & \textbf{0.766}${^*}$ & 0.694 & \underline{0.741} & 0.740 & 0.777 & \textbf{0.721} & 0.709 & 0.657 & \underline{0.715} & 0.709 & 0.699 & 0.710 & 0.705 & 0.745\\

    \bottomrule
    \end{tabular}
    \label{tab:main_results_with_true_validation_data_short_version}
\end{table*}

\subsection{Experimental setup} \label{sec:experimental_setup}
\textbf{Semi-synthetic data.}
We use three widely used benchmark datasets for LTR: Yahoo! Learning to Rank Challenge (C14B)~\cite{chapelle-pmlr11}, MSLR-WEB10k~\cite{qin-arXiv13}, 
and Istella-S~\cite{lucchese-sigir16}. We generate click data using a two-step process following Ai et al.~\cite{ai-sigir18} and Joachims et al.~\cite{joachims-wsdm17}, which is a widely adopted approach in unbiased learning to rank. First, we train a RankSVM~\cite{joachims-kdd06} model on $1\%$ of the training data using true relevance labels to produce initial ranked lists $\pi_q$ for each query $q$ over the remaining $99\%$ training data. An initial ranker is used as it is better than a random ranker while leaving room for improvement. In the second phase, we generate synthetic clicks for each ranked list $\pi_q$ using PBM~\cite{richardson-www07}, where an item is clicked if it is observed ($O^{y}_{q}$) and perceived as relevant ($R^{y}_{q}$). Hence, the click probability can be written as: $Pr(C^{y}_{q} = 1) = Pr(O^{y}_{q}=1|\pi_q) \ Pr(R^{y}_{q}=1|\pi_q)$. Following Joachims et al.~\cite{joachims-wsdm17}, the observation probability can be expressed as: $Pr(O^{y}_{q}=1|\pi_q) = {(\frac{1}{r(y|\pi_q)})}^\eta$, where $\eta$ controls the severity of position bias. If not explicitly mentioned, we set $\eta = 1.0$, which is considered the standard position bias. According to Chapelle et al.~\cite{chapelle-cikm09}, the relevance probability is: $Pr(R^{y}_{q}=1|\pi_q) = \epsilon + (1-\epsilon) \frac{2^{rel(q,y)}-1}{2^{{rel(q,y)}_{max}} - 1}$, where the true relevance for item $y$ with query $q$ is denoted by $rel(q,y) \in [0,4]$. The maximum relevance $rel(q,y)_{max}$ is $4$ in all our datasets. The parameter $\epsilon$ controls click noise, indicating that irrelevant items can still have a chance of being mistakenly perceived as relevant and clicked. By default, we set $\epsilon = 0.0$ unless stated otherwise. 
Generating clicks this way for a full pass over the training data is called one pass. Following Ovaisi et al.~\cite{ovaisi-www20, ovaisi-sigir21}, we use 10 passes over the training dataset to generate clicks, unless stated otherwise. 
We use the clicks generated in this step for all experiments on these benchmark datasets. For RQ7, we also conduct an experiment under stochastic logging policy with randomized interventions, following Joachims et al.~\cite{joachims-wsdm17}. For each query, we apply a randomized intervention on the ranking $\pi_q$, where the item at a designated ``landmark'' position $k$ is swapped with an item randomly selected from rank position $r$, thereby introducing controlled randomness into the item ranking to model stochasticity. We use landmark positions $k \in \{{1, 3, 5, 10\}}$ and swap the item at each landmark with a randomly selected item from another rank position. Unlike a deterministic ranking policy, this enables evaluation when a query–item pair appears at varying positions rather than a fixed position. Then, we generate clicks using the same procedure as discussed earlier. Apart from RQ7, for all other experiments on these benchmark datasets, we use a deterministic logging policy.

\textbf{Real-world data.}
We use the real-world Baidu-ULTR~\cite{zou-neurips22} dataset, which contains billions of query–item pairs collected from Baidu’s search engine, where click data is collected from real users and unbiased relevance labels are available for unbiased evaluation. We provide a summary of the benchmark and real-world datasets in Table~\ref{tab:dataset_description} in Appendix and also a detailed description in Appendix~\ref{datasets_details}.

\textbf{Evaluation metrics.} 
We evaluate using Expected Reciprocal Rank (ERR) and Normalized Discounted Cumulative Gain (NDCG), denoted as $ERR@p$ and $NDCG@p$, where $p$ is the evaluation cutoff.

\textbf{Implementation details.}
LTR literature commonly uses two types of models for ranking: gradient-boosted decision trees (GBDTs) and neural networks (NNs). As a GBDT-based method, we use pairwise LambdaMART~\cite{burges-learning10} as the LTR algorithm in the second stage, which is the most widely used LTR algorithm~\cite{lyzhin-icml23}. We use the LightGBM~\cite{ke-neurips17} implementation for LambdaMART. As an NN-based method, we use a deep neural network (DNN) implemented in~\cite{ai-sigir18} as the LTR algorithm in the second stage. For RQ4, we use debiased clicks estimated from our click debiasing method for hyperparameter tuning. For all other experiments, we use ground-truth relevance from the validation set to tune hyperparameters for both our method and the baselines. The residual transformation is treated as a hyperparameter and tuned using either ground-truth relevance labels or our debiased validation clicks. Following Ai et al.~\cite{ai-sigir18}, we compute statistical significance using the Fisher randomization test ~\cite{smucker-cikm07} with $\bm{p} < 0.05$. We provide more detailed implementation details, including hyperparameter settings, first-stage model specification, and training parameters, in Appendix~\ref{more_implementation_details}.

\textbf{Baselines.} We compare CFC with position bias correction baselines, including PIJD~\cite{ovaisi-sigir21}, PairD~\cite{hu-www19}, UPE~\cite{luo-sigir24}, MULTR~\cite{luo-wsdm23}, PAL~\cite{guo-recsys19}, REM~\cite{wang-wsdm18}, and DLA~\cite{ai-sigir18}. These baselines represent the main methodological families of state-of-the-art position bias correction in unbiased learning to rank, including IPW-based methods (e.g., UPE, REM, PairD, and DLA), click-model–based approaches (PAL), econometric methods (PIJD), and DR IPW-based methods (MULTR). Since CFC method is implemented with the LambdaMART and the DNN rankers, we include comparisons with the LambdaMART and DNN ranking algorithms that do not correct for any bias. The Oracle method is trained with human-annotated labels, representing the upper-bound performance of any ranker. For PIJD, we use both LambdaMART and DNN as the ranker, as PIJD is flexible to integrate with both of these rankers. For PairD, we use LambdaMART, as PairD is coupled with LambdaMART in the original implementation. We use DNN ranker for UPE, MULTR, PAL, REM, and DLA, as these methods are implemented with DNNs. On the other hand, our method is flexible enough to integrate with both of the rankers. We provide a detailed description of the baselines in Appendix~\ref{baselines}.

\textbf{Heteroskedasticity test on first-stage residuals.} \label{sec:heteroskedasticity}
To validate heteroskedasticity in the first-stage residuals, we apply the nonparametric 
Fligner--Killeen test~\cite{fligner-jasa76} to the Yahoo, MSLR-WEB10K, and Istella-S datasets. Across all datasets, the test is statistically significant ($\bm{p} \approx 10^{-6}$--$10^{-5} < 0.05$), confirming heteroskedasticity and satisfying the key assumption required by the Lewbel method for constructing control function terms. We report statistics of the first-stage residuals under different production ranker settings in Table~\ref{tab:prod_residual_variations} and provide a detailed analysis in Appendix~\ref{more_ablations}.

\begin{table*} [b]
    \vspace{-0.5em}
  \fontsize{7.8}{7}\selectfont
  \captionsetup{justification=raggedright, width=1.0\textwidth}
  \caption{A comparison of the overall performance of CFC and other baselines with real-world Baidu-ULTR dataset. The best-performing method for each ranker is shown in bold and the second-best is underlined.}
  \vspace{-0.05em}
  \centering
  \begin{tabular}{ccccccccccccc}
    \toprule
    \multirow{2}{*}{Metrics} & \multicolumn{4}{c}{LambdaMART} & \multicolumn{7}{c}{DNN} \\ 
    \cmidrule(r){2-5} \cmidrule(r){6-13}
    & CFC & PIJD & PairD & LambdaMART 
    & CFC & PIJD & UPE & MULTR & PAL & REM & DLA & DNN \\
    \midrule

    ERR@10 & \textbf{0.275} & 0.234 & 0.234 & \underline{0.238} 
           & \textbf{0.291} & 0.247 & 0.222 & 0.222 & 0.262 & \underline{0.279} & 0.213 & 0.223 \\

    NDCG@10 & \textbf{0.525} & 0.442 & 0.448 & \underline{0.454} 
            & \textbf{0.553} & 0.475 & 0.442 & 0.437 & 0.502 & \underline{0.532} & 0.434 & 0.415 \\

    \bottomrule
  \end{tabular}
  \label{tab:baidu_dataset_results}
\end{table*}

\begin{table*} [b]
  \setlength{\tabcolsep}{4.8pt} 
  \fontsize{7.8}{7}\selectfont
  \captionsetup{justification=raggedright, width=1.0\textwidth}
    \caption{A comparison of the overall performance of CFC and other baselines using click labels for hyperparameter tuning. The best-performing method for each ranker, excluding Oracle, is shown in bold, and the second-best is underlined. * denotes a statistically significant improvement of the best-performing method over the best baseline, considering all LambdaMART and DNN rankers, with ${\bm{p < 0.05}}$.}
  \vspace{-0.05em}
  \centering
    \begin{tabular}{cccccccccccccccc}
    \toprule
    \multirow{2}{*}{Datasets} & \multirow{2}{*}{Metrics} & \multicolumn{5}{c} {LambdaMART} & \multicolumn{9}{c}{DNN} \\ \cmidrule(r){3-7} \cmidrule(r){8-16}
        & & CFC & PIJD & PairD & LambdaMART & Oracle & CFC & PIJD & UPE & MULTR & PAL & REM & DLA & DNN & Oracle \\
     \midrule

    \multirow{2}[0]{*}{Yahoo}
    & ERR@10 & \textbf{0.458}${^*}$ & \underline{0.452} & \underline{0.452} & 0.446 & 0.474 & \textbf{0.453} & 0.447 & 0.437 & \underline{0.452} & 0.441 & 0.444 & 0.433 & 0.441 & 0.464 \\
    & NDCG@10 & \textbf{0.792}${^*}$ & 0.780 & 0.785 & \underline{0.789} & 0.817 & \textbf{0.791} & 0.785 & 0.772 & \underline{0.789} & 0.782 & 0.768 & 0.777 & 0.781 & 0.805\\
    \midrule

    \multirow{2}[0]{*}{\makecell{MSLR-\\WEB10k}}
    & ERR@10 & \textbf{0.321}${^*}$ & 0.275 & \underline{0.285} & 0.268 & 0.342 & \underline{0.266} & 0.259 & 0.232 & \textbf{0.291} & 0.253 & 0.241 & 0.256 & 0.253 & 0.325 \\
    & NDCG@10 & \textbf{0.490}${^*}$ & 0.438 & \underline{0.463} & 0.446 & 0.498 & \textbf{0.457} & 0.448 & 0.362 & \underline{0.451} & 0.436 & 0.419 & 0.447 & 0.446 & 0.476 \\
    \midrule

    \multirow{2}[0]{*}{Istella-S}
    & ERR@10 & \textbf{0.715}${^*}$ & 0.680 & 0.689 & \underline{0.690} & 0.737 & \textbf{0.691} & 0.672 & 0.671 & 0.675 & 0.669 & 0.653 & \underline{0.676} & 0.669 & 0.702 \\
    & NDCG@10 & \textbf{0.762}${^*}$ & 0.692 & \underline{0.739} & 0.733 & 0.777 & \textbf{0.717} & \underline{0.706} & 0.690 & 0.691 & 0.708 & 0.685 & \underline{0.706} & 0.705 & 0.745 \\

    \bottomrule
    \end{tabular}
    \label{tab:main_results_with_biased_clicks_short_version}
\end{table*}

\subsection{Results}
In this section, we show the results of our proposed CFC method.

\textbf{Performance comparison between CFC and other methods (RQ1).} \label{sec:performance_true}
For RQ1, we compare CFC with other baselines in simulation experiments. Results are shown in Table~\ref{tab:main_results_with_true_validation_data_short_version}. Across all three benchmark datasets, CFC outperforms all corresponding baselines on both $ERR@10$ and $NDCG@10$, using either LambdaMART or DNN rankers, demonstrating the effectiveness of CFC. Because our method outperforms the corresponding baselines for both rankers, this also demonstrates the generalizability of our method across different LTR algorithms for position bias correction.
Furthermore, clicks in the MSLR-WEB10k and Istella-S datasets are sparser than in Yahoo, with click densities of approximately $0.36\%$ and $1.10\%$, respectively, compared to $2.72\%$ in Yahoo under standard position bias. CFC method surpasses the baselines, showing its robustness in scenarios with such highly sparse click data. On the MSLR-WEB10K and Istella-S, the PDF and IMR transformations assuming normality yield the best ranking performance, likely because the residuals from $m(\cdot)$ in these datasets approximately follow a normal distribution. In contrast, for the Yahoo dataset, where the residuals do not follow an approximately normal distribution, the min-max transformation performs comparatively better. This highlights that CFC method can automatically find the best residual transformation to achieve optimal ranking performance. Overall, CFC with LambdaMART performs best across both LambdaMART and DNN-based methods, suggesting that GBDT-based rankers are more effective for LTR than neural network–based rankers, consistent with~\cite{hager-sigir24}.  We also evaluate performance for $p \in \{{1, 3, 5\}}$ on both ERR and NDCG, as shown in Table~\ref{tab:main_results_with_true_validation_data_short_version_detailed} in the Appendix. Across all values of $p$, CFC outperforms the corresponding baselines for both rankers.

\begin{figure}
  \centering
  \captionsetup{justification=raggedright, margin=0cm}
  \captionsetup[subfigure]{labelfont=normalfont, textfont=normalfont}

  \begin{subfigure}{0.48\textwidth}
    \includegraphics[width=\linewidth]{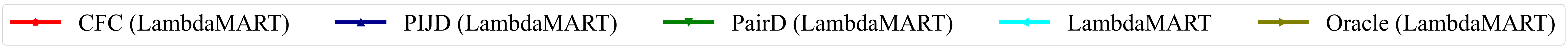}
  \end{subfigure}
  \begin{subfigure}{0.15\textwidth}
    \includegraphics[width=\linewidth]{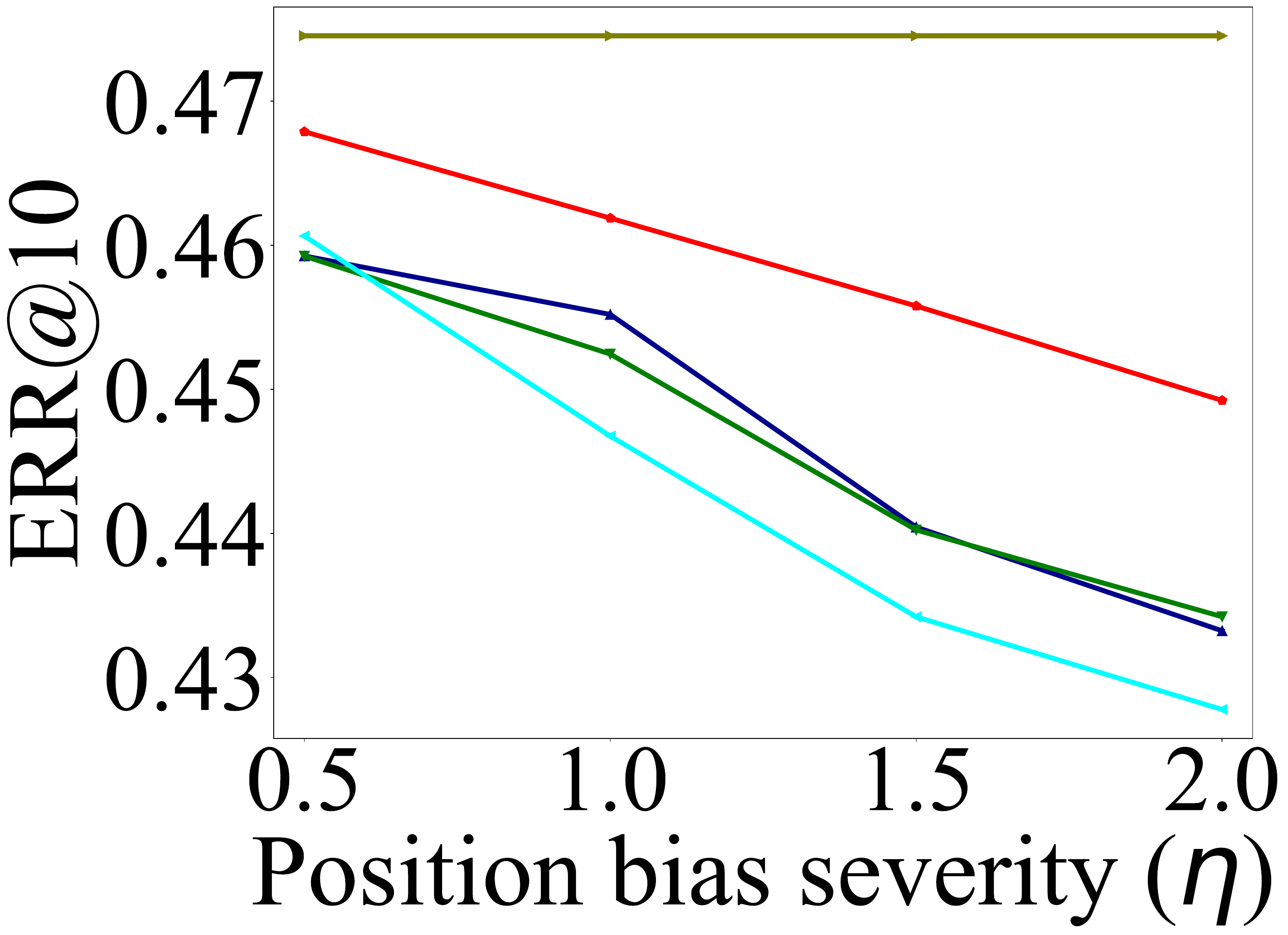}
    \caption{Yahoo}
  \end{subfigure}
  \begin{subfigure}{0.15\textwidth}
    \includegraphics[width=\linewidth]{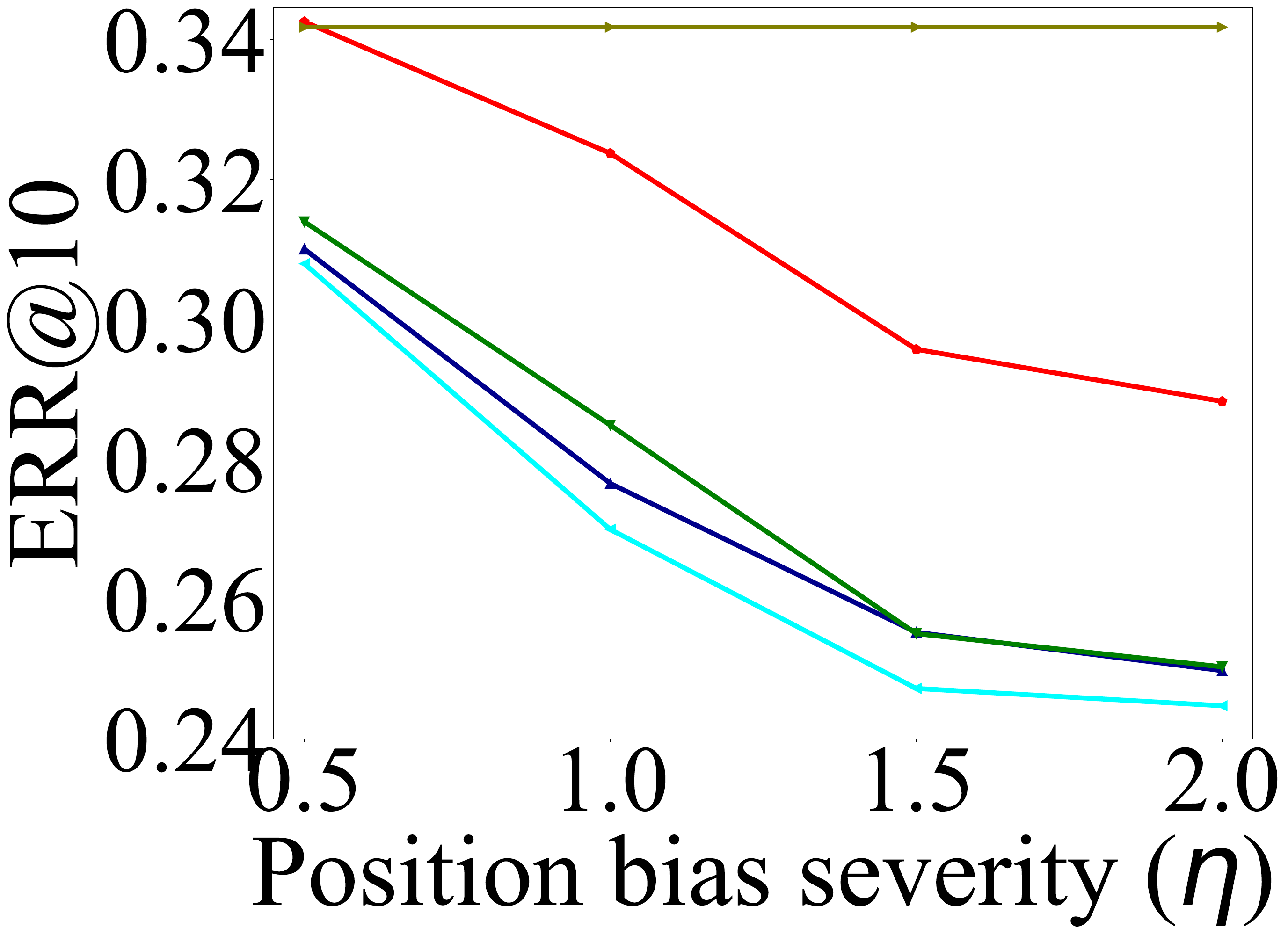}
    \caption{MSLR-WEB10k}
  \end{subfigure}
  \begin{subfigure}{0.15\textwidth}
    \includegraphics[width=\linewidth]{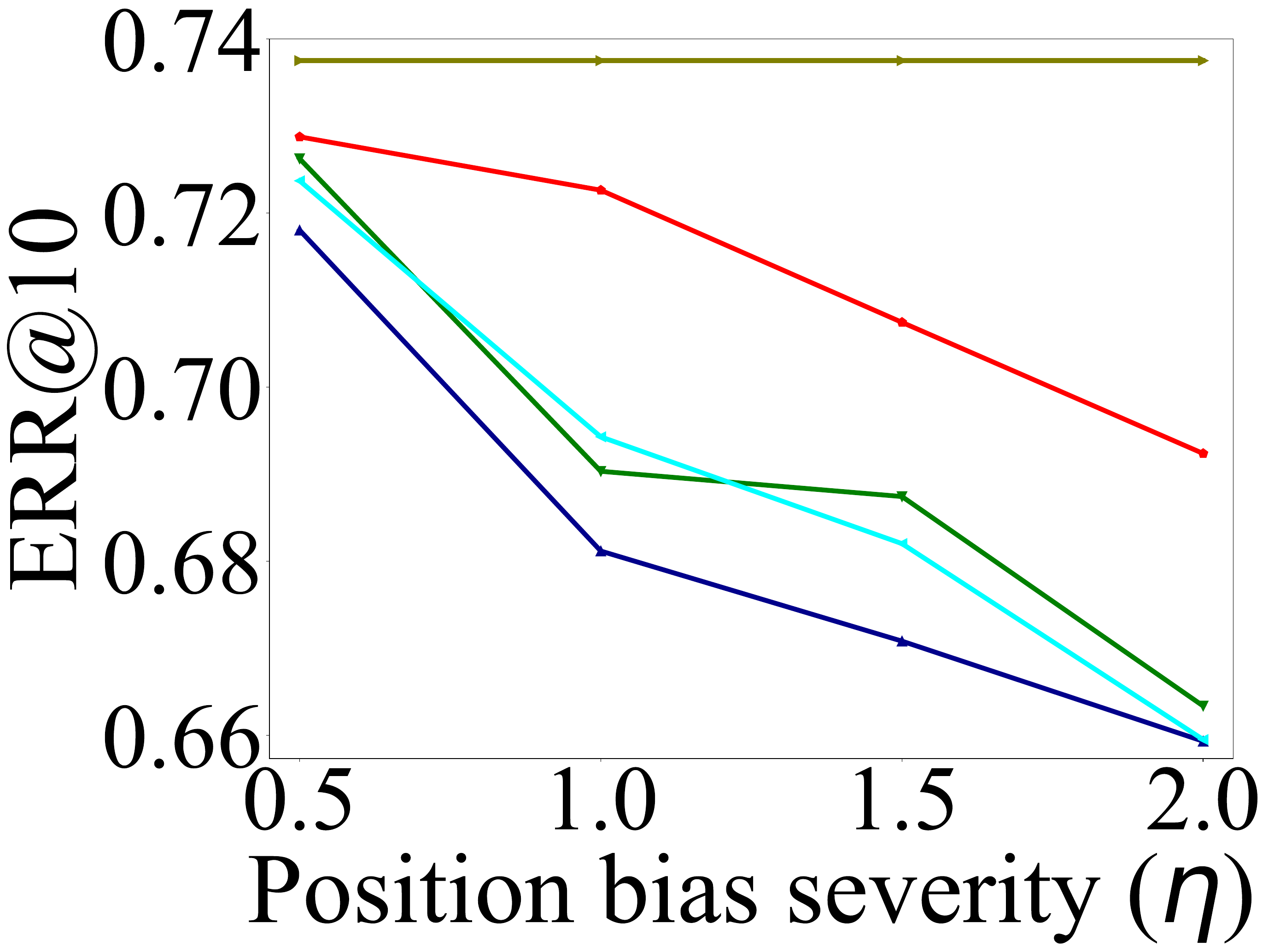}
    \caption{Istella-S}
  \end{subfigure}


  \begin{subfigure}{0.48\textwidth}
    \includegraphics[width=\linewidth]{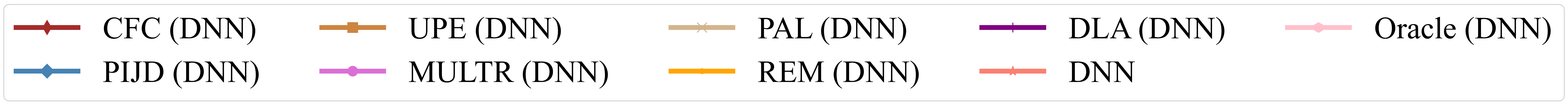}
  \end{subfigure}
  \begin{subfigure}{0.15 \textwidth}
    \includegraphics[width=\linewidth]{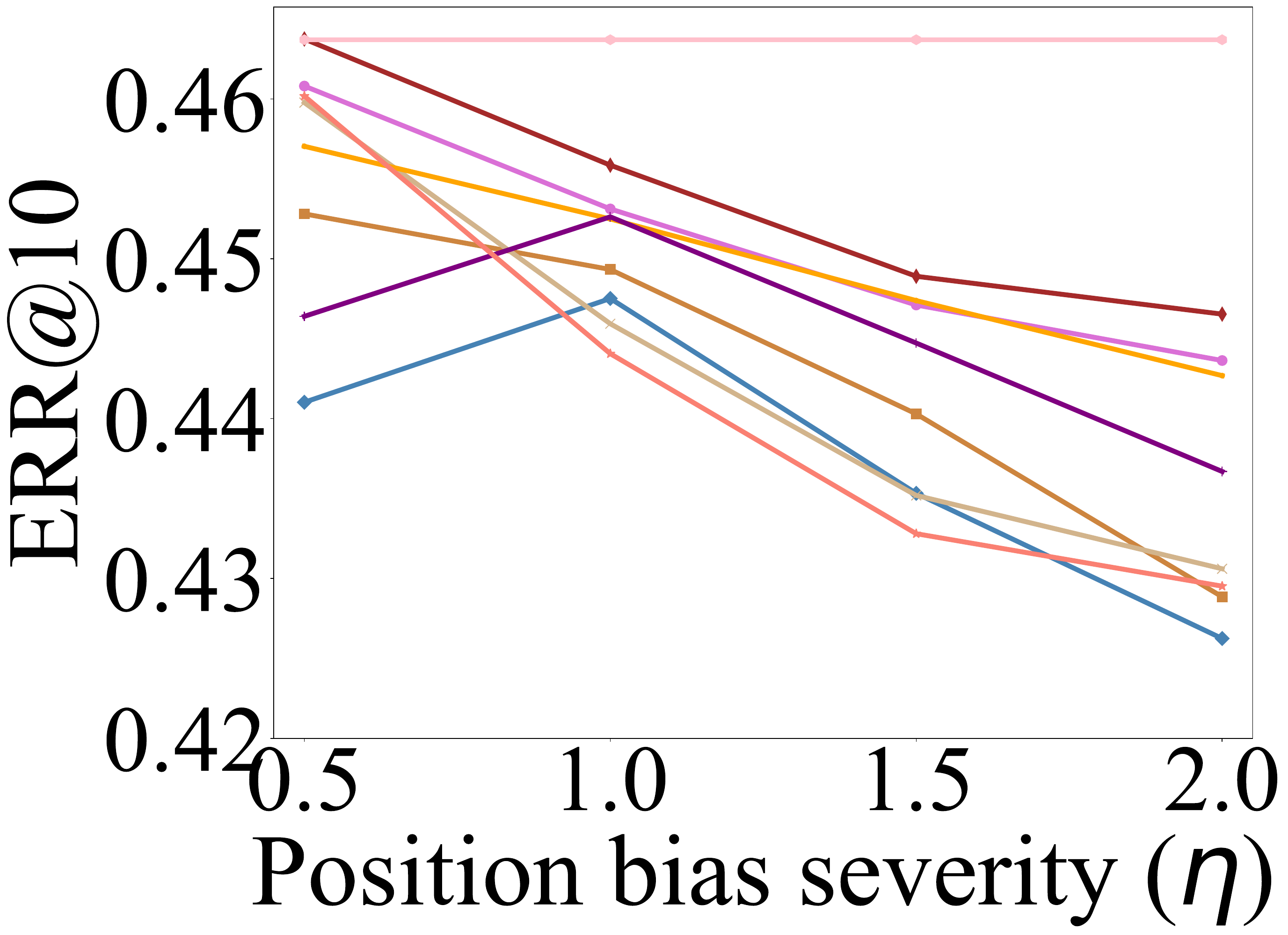}
    \caption{Yahoo}
  \end{subfigure}
  \begin{subfigure}{0.15 \textwidth}
    \includegraphics[width=\linewidth]{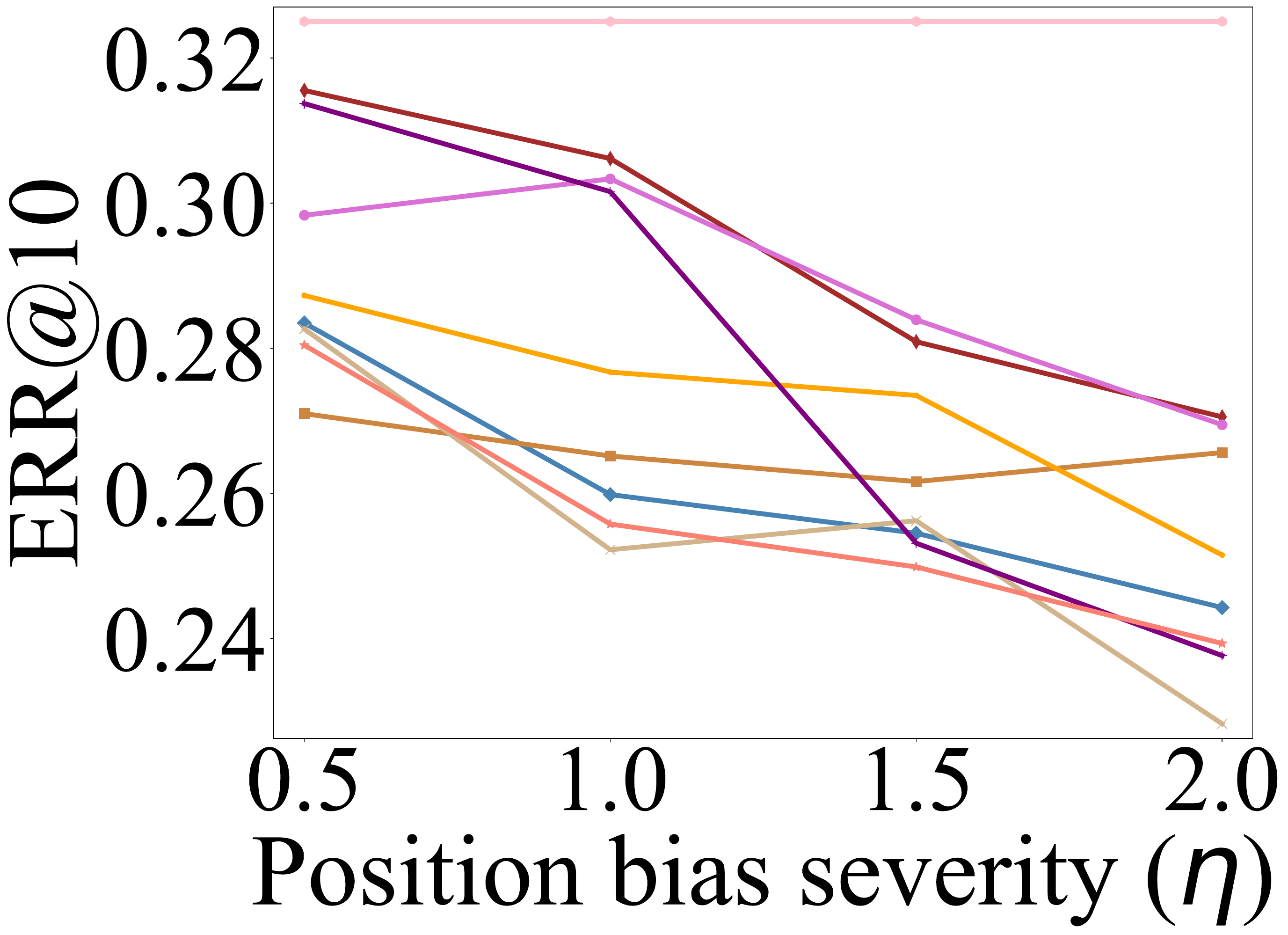}
    \caption{MSLR-WEB10k}
  \end{subfigure}
  \begin{subfigure}{0.15 \textwidth}
    \includegraphics[width=\linewidth]{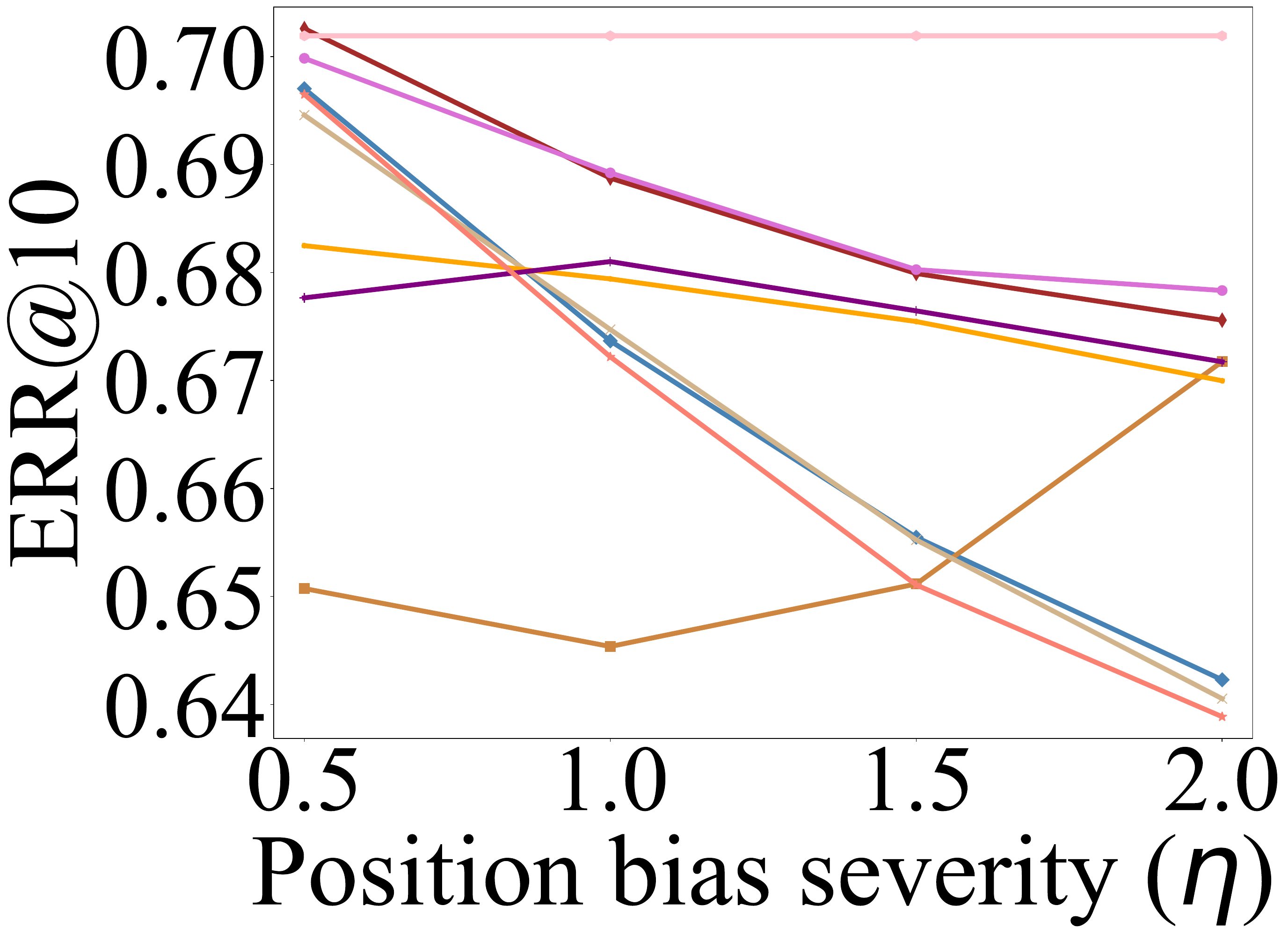}
    \caption{Istella-S}
  \end{subfigure}
  \vspace{-0.25cm}
    
  \captionsetup{width=0.48\textwidth}
  \caption{The performance of different methods with varying levels of position bias severity.\label{fig:bias_severity-all}}
\end{figure}

\textbf{The effect of bias severity (RQ2).}
We evaluate CFC by varying the severity of position bias (\( \eta \in \{0.5, 1.0, 1.5, 2.0\} \)) in click generation. Figure~\ref{fig:bias_severity-all} shows the $ERR@10$ results with LambdaMART and DNN as rankers. Across all datasets and bias severities, CFC with LambdaMART outperforms all baselines. As \( \eta \) increases, the performance gap between Oracle and other methods increases due to increased sparsity in clicks. Nevertheless, our method is more robust than baselines across bias severity levels. CFC with DNN also performs best among DNN-based baselines on Yahoo and MSLR-WEB10K for all \( \eta \), except for \( \eta = 1.5 \) on MSLR-WEB10K. On the Istella-S dataset, CFC with DNN outperforms the baselines for \( \eta = 0.0 \) and \( 2.0 \), and performs comparably to MULTR for 
\( \eta = 1.0 \) and \( 1.5 \) among DNN-based methods. MULTR sometimes  performs comparably to CFC with a DNN ranker, likely because MULTR is optimized to work with a DNN ranker. However, unlike CFC, which is a general-purpose position bias correction method that can be applied to any LTR algorithm to correct for position bias, MULTR cannot generalize to other rankers without substantial effort.

\textbf{Evaluation on real-world industrial search engine (RQ3).}
We evaluate CFC on the real-world Baidu-ULTR dataset, which contains click logs collected from Baidu’s search engine. We use these click logs for training and evaluate all methods using the unbiased ground-truth relevance labels provided in the dataset. We show the results in Table~\ref{tab:baidu_dataset_results}. Across both $ERR@10$ and $NDCG@10$, CFC outperforms all corresponding baselines under both  rankers, highlighting the effectiveness of CFC in large-scale commercial search settings. Many existing unbiased learning to rank methods perform well on benchmark datasets where the production ranker is relatively weak (e.g., does a poor job of ranking items based on their relevance)~\cite{luo-sigir24}. However, these methods often fail to generalize to real-world settings where the production ranker is highly optimized, as in typical industrial LTR systems~\cite{hager-sigir24}. In such cases, these methods struggle due to the stronger coupling between position generated by production ranker and underlying relevance~\cite{luo-sigir24}. CFC remains effective on the Baidu dataset, where the click logs are generated by a strong production ranker~\cite{hager-sigir24}, demonstrating that CFC can mitigate position bias even under strong logging policies. We do not include Oracle performance because the training data does not contain ground-truth relevance. We also evaluate performance for $p \in \{{1, 3, 5\}}$ on both ERR and NDCG, as shown in Table~\ref{tab:baidu_dataset_results_detailed} in the Appendix. Across all values of $p$, CFC outperforms the corresponding baselines for both rankers.

\textbf{Performance comparison without access to true relevance in validation data (RQ4).} 
In response to RQ4, we assume true relevance labels are unavailable in the validation dataset and use click data for hyperparameter tuning. For CFC, we use debiased clicks estimated from click data using our debiasing click method.
Since the baselines do not have a mechanism to debias the click data, we use the click data for hyperparameter tuning in the baselines. Results are shown in Table~\ref{tab:main_results_with_biased_clicks_short_version}. Across all datasets, CFC achieves the best performance among all corresponding baselines for both $ERR@10$ and $NDCG@10$ metrics when using either LambdaMART or DNN rankers. However, only for $ERR@10$ on the MSLR-WEB10k dataset with the DNN ranker, CFC is the second-best method, while MULTR performs best. This may be attributed to MULTR being specifically tailored to the DNN ranker. Nevertheless, CFC with LambdaMART outperforms all LambdaMART- and DNN-based baselines for both metrics. This highlights the effectiveness of our approach when actual relevance labels for validation data are unavailable for selecting the best models via hyperparameter tuning.

\textbf{RQ5-RQ7.}
In RQ5, we evaluate the performance of the CFC method with increasing amounts of click data. Across different click volumes, our method performs better than the baselines for both LambdaMART and DNN rankers in all datasets (Figure~\ref{fig:passes_variations-all} in Appendix). As the number of clicks increases, performance improves and approaches the Oracle, benefiting from a bias–variance tradeoff with reduced bias and variance. To assess robustness to noisy clicks, we vary the probability of clicks on irrelevant items in RQ6. Our method remains robust and outperforms baselines across different noise levels for both rankers (Figure~\ref{fig:noisy_clicks} in Appendix). In RQ7, we evaluate the effectiveness of CFC under a stochastic logging policy. CFC outperforms corresponding baselines for both LambdaMART and DNN rankers across all datasets (Table~\ref{tab:main_results_with_true_validation_data_short_version_stochastic_detailed} in Appendix). The stochastic logging policy introduces variation in item rankings, leading to greater variation in ranking residuals, which further improves performance compared to the deterministic setting in RQ1. Although baselines, particularly IPW-based methods, also show slight improvements due to more accurate propensity estimation under stochastic logging, CFC achieves a  consistent performance improvement over baselines. We provide a detailed analysis of RQ5, RQ6, and RQ7 in Subsection~\ref{sec:detailed_exp_research_qstn} in the Appendix.

\textbf{Ablation Studies.} 
We provide several ablation studies to analyze the proposed CFC method. We vary the production ranker (RankSVM) by training it with different fractions of true relevance data (1\%, 10\%, 20\%, 30\%, and 100\%) to generate rankings. The performance of CFC remains stable when using different production rankers, demonstrating robustness to the logging policy specification (Table~\ref{tab:prod_variations} in Appendix). Even with stronger rankers (i.e., trained with more data, such as 100\%), performance remains comparable, and CFC mitigates bias under both weak and strong production rankers. We conduct another study on a simplified setting that uses only first-stage residuals in the second-stage click model, instead of using feature interaction terms. While the CFC approach generally performs better, the residual-only formulation remains competitive and also provides a meaningful degree of bias correction in terms of ranking performance (Table~\ref{tab:cfc_cfc_s_diff} in Appendix). We also use different machine learning models to estimate $m(\cdot)$ (Ridge, Logistic Regression, SVM, and XGBoost) and find that CFC is robust to the choice of first-stage model, with all models yielding comparable ranking performance when incorporating residuals into the second stage (Table~\ref{tab:cf_variations} in Appendix).
In addition, the first-stage models are computationally efficient.
We also analyze first-stage residuals estimated from $m(\cdot)$ under production rankers trained with varying amounts of labeled data. Stronger rankers generally yield smaller mean residuals, indicating that they explain more of the feature-dependent variation in item rankings (Table~\ref{tab:prod_residual_variations} in Appendix). Statistically significant heteroskedasticity persists across all production ranker settings ($\bm{p < 0.05}$), supporting the validity of CFC under different production rankers. Moreover, we also highlight the importance of using debiased clicks versus raw biased clicks for hyperparameter tuning. Results show that using debiased clicks leads to better performance than relying on biased clicks, underscoring the importance of debiasing the validation clicks (Table~\ref{tab:cfc-with-bc} in Appendix). We provide a detailed description and analysis of all these studies in the ablation studies in Appendix~\ref{more_ablations}.
\vspace{-1.0em}

\section{Conclusion}



We introduce a new control function–based method to correct for position bias in LTR systems. We first estimate the historical ranking process that generated the observed user feedback using query–item features, and then use the resulting ranking residuals and feature interaction terms, as control functions in a second-stage model to account for position bias. Our method does not require propensity estimation or modifications to the ranker’s objective function, making it applicable to any state-of-the-art LTR method for position bias correction. Another contribution of this study is a residual-based click debiasing approach for hyperparameter tuning in the absence of annotated validation data. Evaluations show that CFC outperforms state-of-the-art methods on both benchmark and large-scale industrial datasets and reduces position bias under both deterministic and stochastic logging policies. CFC is also robust to varying levels of position bias severity, noisy clicks, and both weak and strong production rankers that generated the click data.
The CFC framework is flexible and can be extended to model additional sources of bias, such as trust bias, selection bias, or click noise, which we leave for future work. Moreover,  
complete separation between bias and feature-dependent relevance may not always be achievable because the control function terms may interact with the features during training. Consequently, setting the control function terms to zero at inference may also remove part of the learned feature–control function interactions. A promising direction for future work is to develop architectures and training objectives that more effectively disentangle bias-related and relevance-related components while preserving the flexibility of highly nonlinear models. 


\begin{acks}
This research was partly supported by the National Science Foundation under Award No. 2217023.
\end{acks}

\bibliographystyle{ACM-Reference-Format}
\balance
\bibliography{main}


\clearpage
\appendix

\section{More Analysis of the CFC Method}

\subsection{CFC algorithm} \label{cfc_algorithm}
The complete CFC algorithm is provided below.
\begin{algorithm}
    \caption{CFC method for position bias correction}
    \label{alg:cfc_bias_correction}
    \begin{algorithmic}[1]

        \Require Features set $\bm{X}$, item clicks indicators $C$, and item rankings $\pi$
        \Ensure An unbiased ranking model

        \State Train a ranking model $m()$ using Eq.~\eqref{eq:control_function}
        \State Compute the residuals ($\hat{\epsilon}^{y}_{q, m}$) for each query-item pair

        \State $T_1 \leftarrow$ apply min-max normalization of residuals
        \State $T_2 \leftarrow$ apply parametric PDF transformation of residuals
        \State $T_3 \leftarrow$ apply parametric IMR transformation of residuals
        \State $T_4 \leftarrow$ apply nonparametric hazard ratio transformation of residuals

        \State Initialize the minimum loss: $\text{min\_loss} \leftarrow \infty$
        \State Initialize the best model: $\text{best\_model} \leftarrow NULL$
        
        \For{each transformation $T_i$ (from $T_1$ to $T_4$)}
            \State Train a ranking model with both clicked and unclicked items and the transformed residuals $T_i$ using Eq.~\eqref{eq:unbiased_c_lewbel}
            \State Estimate the ranking loss $\hat{\bm{\mathcal{L}}}$ on validation data for the trained model $g(\cdot)$
            \If{$\hat{\bm{\mathcal{L}}} < \text{min\_loss}$}
                \State $\text{min\_loss} \leftarrow \hat{\bm{\mathcal{L}}}$
                \State $\text{best\_model} \leftarrow g(\cdot)$
            \EndIf
        \EndFor

        \State \Return best\_model

    \end{algorithmic}
\end{algorithm}

\subsection{An illustrative example of the CFC method} \label{sec:cfc_example}
To illustrate how our control function approach corrects for position bias, we provide a simple example with four items shown for a query. We demonstrate how residuals are computed from the item ranking process and used in the second stage click equation.

Let us consider a user searching for a query $q$, where the system previously ranked four items. Each item has three features: length, TF-IDF score, and recency. The rank of each item $r(y\mid \pi_q)$ and their features $\bm{x}_{q}^{y}$ are shown in Table~\ref{tab:example_cfc}.
\begin{table} [b]
  \fontsize{7.8}{7}\selectfont
  \captionsetup{justification=raggedright, width=0.49\textwidth}
  \caption{Features and rankings of four items associated with query $q$.} 
  \centering
  \begin{tabular}{ccccc}
    \toprule
    item & Length & TF-IDF & Recency (days) & Rank ($r(y|\pi_q)$) \\
    \midrule
    A & 120 & 0.5 & 2 & 3 \\
    B & 90  & 0.8 & 1 & 4 \\
    C & 100 & 0.6 & 4 & 1 \\
    D & 110 & 0.4 & 3 & 2 \\
    \bottomrule
    \end{tabular}
    \label{tab:example_cfc}
\end{table}
So the feature vectors $\bm{x}_{q}^{y}$ of each item $y$ are:
\begin{align*}
    \bm{x}_A = [120, 0.5, 2],\quad
    \bm{x}_B = [90, 0.8, 1],\\
    \bm{x}_C = [100, 0.6, 4],\quad
    \bm{x}_D = [110, 0.4, 3]
\end{align*}

We, at first, train the first-stage ranking model $m(\cdot)$ using features $\bm{x}_{q}^{y}$ as inputs and item ranks $r(y \mid \pi_q)$ as target outputs. Let the first-stage item ranking model $m(\cdot)$ predict the following ranks:
\[
\hat{r}(A|\pi_q) = 2,\quad
\hat{r}(B|\pi_q) = 6,\quad
\hat{r}(C|\pi_q) = -2,\quad
\hat{r}(D|\pi_q) = 3
\]
Next, we calculate the residuals for each item as $\hat{\epsilon}^{y}_{q,m} = r(y|\pi_q) - \hat{r}(y|\pi_q)$, as follows:
\begin{align*}
\hat{\epsilon}^{A}_{q,m} = 3 - 2 = +1,\quad
\hat{\epsilon}^{B}_{q,m} = 4 - 6 = -2,\\
\hat{\epsilon}^{C}_{q,m} = 1 - (-2) = +3,\quad
\hat{\epsilon}^{D}_{q,m} = 2 - 3 = -1
\end{align*}
Now, we apply transformations to the estimated residuals. For simplicity, let us apply only one type of transformation, such as min-max normalization, to transform the residuals. Thus, we get:
\begin{align*}
    T(\hat{\epsilon}^{A}_{q,m}) = \frac{1 + 2}{5} = 0.6,\quad
    T(\hat{\epsilon}^{B}_{q,m}) = \frac{-2 + 2}{5} = 0.0,\\
    T(\hat{\epsilon}^{C}_{q,m}) = \frac{3 + 2}{5} = 1.0,\quad
    T(\hat{\epsilon}^{D}_{q,m}) = \frac{-1 + 2}{5} = 0.2
\end{align*}

Next, we incorporate the residual in the control function $\Phi$ using: $
\Phi\left(\bm{x}_{q}^{y}, T(\hat{\epsilon}^{y}_{q,m})\right) = (\bm{x}_{q}^{y} - \bar{\bm{x}}) \odot T(\hat{\epsilon}^{y}_{q,m}),
$ where the feature mean in each dimension calculated as: $
\bar{\bm{x}} = \frac{1}{4}(\bm{x}_A + \bm{x}_B + \bm{x}_C + \bm{x}_D) = [105, 0.575, 2.5].$ Now, by interacting each demeaned feature with the ranking residual, we obtain:
\begin{align*}
\Phi_A &= ([120, 0.5, 2] - [105, 0.575, 2.5]) \cdot 0.6 = [9.0, -0.045, -0.3] \\
\Phi_B &= ([90, 0.8, 1] - [105, 0.575, 2.5]) \cdot 0.0 = [0, 0, 0] \\
\Phi_C &= ([100, 0.6, 4] - [105, 0.575, 2.5]) \cdot 1.0 = [-5, 0.025, 1.5] \\
\Phi_D &= ([110, 0.4, 3] - [105, 0.575, 2.5]) \cdot 0.2 = [1.0, -0.035, 0.1]
\end{align*}

Now, the second stage final click model is trained using $\phi(\cdot)$ and query-item features $\bm{x}_{q}^{y}$ as input features and clicks as target outputs. At test time, when predicting items for new queries, we use \( g(\bm{x}_q^y, 0) \) to obtain unbiased click probabilities for items, which are then used to rank items for that query. This example illustrates how residuals are estimated from the first-stage ranking model $m(\cdot)$, transformed, and integrated into the second stage click model as control function terms to obtain unbiased click probabilities.

\section{More Experimental Details} \label{experimental_details}
\subsection{Datasets} \label{datasets_details}
We now provide detailed descriptions of the three benchmark datasets used for evaluation: Learning to Rank Challenge (C14B)\footnote{\label{yahoo_link}\url{https://webscope.sandbox.yahoo.com/catalog.php?datatype=c}}~\cite{chapelle-pmlr11}, MSLR-WEB10K\footnote{\label{mslr_link}\url{https://www.microsoft.com/en-us/research/project/mslr}}~\cite{qin-arXiv13}, and Istella-S\footnote{\label{istella_link}\url{http://quickrank.isti.cnr.it/istella-dataset}}~\cite{lucchese-sigir16}. Each benchmark dataset contains human-annotated relevance labels for each query-item pair $\in [0,4]$. 
For the real-world Baidu-ULTR dataset\footnote{\label{baidu_link}\url{https://github.com/ChuXiaokai/baidu_ultr_dataset/}}~\cite{zou-neurips22}, the dataset directly provides click data and ground-truth relevance labels for each query-item pair in the range $[0,4]$ for evaluation.
\begin{itemize}[leftmargin=10pt, nosep]
    \item \textbf{Yahoo! Learning to Rank Challenge} dataset includes $19,944$ training queries with $473,134$ items. It also contains $2,994$ validation queries with $71,083$ items, and $6,983$ test queries with $165,660$ items. On average, there are $24$ items per query and each item has a $700$-dimensional feature vector.
    \item \textbf{MSLR-WEB10k} dataset has five folds. We use fold-1 for our evaluations from the MSLR-WEB10k dataset. Fold-1 consists of $6,000$ training queries with $723,412$ items, $2,000$ validation queries with $235,260$ items, and $2,000$ test queries with $241,521$ items. Each item has a $136$-dimensional feature vector with an average of $121$ items per query.
    \item \textbf{Istella-S} dataset consists of $19,245$ training queries with $2,043,304$ items, $7,211$ validation queries with $684,076$ items, and $6,562$ test queries with $681,250$ items. There is an average of $103$ items per query where each item has a $220$-dimensional feature vector.
    \item \textbf{Baidu-ULTR} is a large-scale industrial dataset containing billions of query-item pairs and is collected from click logs of the Baidu search engine. It also provides unbiased ground truth relevance labels for unbiased evaluation. We extract the dataset using the Hugging Face API\footnote{\label{baidu_huggigface}\url{https://huggingface.co/datasets/philipphager/baidu-ultr_uva-mlm-ctr}}. Baidu-ULTR dataset consists of $1{,}779{,}017$ training queries with $14{,}526{,}276$ items, $136{,}513$ validation queries with $3{,}455$ items, and $261{,}009$ test queries with $18$ items. Each item is represented by an $18$-dimensional LTR feature vector.
\end{itemize}
\begin{table}
  \vspace{-1em}
  \setlength{\tabcolsep}{1.9pt} 
  \fontsize{7.2}{8}\selectfont
  \captionsetup{justification=raggedright, width=1.0\textwidth}
  \caption{Dataset descriptions.}
  \vspace{-1em}
  \centering
  \begin{tabular}{cccccccc}
    \toprule
    Dataset & \makecell{Train \\ queries} & \makecell{Train \\ docs.} & \makecell{Valid \\ queries} & \makecell{Valid \\ docs.} & \makecell{Test \\ queries} & \makecell{Test \\ docs.} & 
    \makecell{LTR \\ feature} \\
    \midrule
    Yahoo (Set1) & 19,944 & 473,134 & 2,994 & 71,083 & 6,983 & 165,660 & 
    700 \\
    \makecell{MSLR-WEB10k} & 6,000 & 723,412 & 2,000 & 235,260 & 2,000 & 241,521 & 
    136 \\
    Istella-S & 19,245 & 2,043,304 & 7,211 & 684,076 & 6,562 & 681,250 &
    220 \\
    Baidu-ULTR & 1,779,017 & 14,526,276 & 1,727 & 136,513 & 3,455 & 261,009 &
    18 \\ 
    \bottomrule
  \end{tabular}
  \label{tab:dataset_description}
\end{table}
We present a summary of the descriptions of these three datasets in Table \ref{tab:dataset_description}.

\begin{table*}
  \setlength{\tabcolsep}{4.8pt} 
  \fontsize{7.8}{7}\selectfont
  \captionsetup{justification=raggedright, width=1.0\textwidth}
  \caption{A comparison of the overall performance of CFC and other baselines. The best-performing method for each ranker, excluding Oracle, is shown in bold, and the second-best is underlined. * denotes a statistically significant improvement of the best-performing method over the best baseline, considering all LambdaMART and DNN rankers, with ${\bm{p < 0.05}}$.}
  \vspace{-1em}
  \centering
  \begin{tabular}{cccccccccccccccc}
  \toprule
  \multirow{2}{*}{Datasets} & \multirow{2}{*}{Metrics} & \multicolumn{5}{c}{LambdaMART} & \multicolumn{7}{c}{DNN} \\
  \cmidrule(r){3-7} \cmidrule(r){8-16}
   & & CFC & PIJD & PairD & LambdaMART & Oracle & CFC & PIJD & UPE & MULTR & PAL & REM & DLA & DNN & Oracle \\
  \midrule

\multirow{7}{*}{Yahoo}
& ERR@1 & \textbf{0.345}${^*}$ & \underline{0.339} & 0.335 & 0.325 & 0.359 
        & \textbf{0.340} & 0.326 & 0.333 & \underline{0.339} & 0.324 & 0.338 & \underline{0.339} & 0.321 & 0.346 \\

& ERR@3 & \textbf{0.425}${^*}$ & \underline{0.416} & 0.414 & 0.408 & 0.438 
        & \textbf{0.419} & 0.409 & 0.411 & \underline{0.417} & 0.407 & 0.416 & 0.414 & 0.405 & 0.426 \\

& ERR@5 & \textbf{0.446}${^*}$ & \underline{0.437} & 0.436 & 0.430 & 0.460 
        & \textbf{0.441} & 0.432 & 0.433 & \underline{0.439} & 0.430 & 0.437 & 0.437 & 0.428 & 0.449 \\
\cmidrule{2-16}

& NDCG@1 & \textbf{0.752}${^*}$ & 0.734 & \underline{0.735} & 0.733 & 0.778 
        & \textbf{0.740} & 0.730 & 0.712 & \underline{0.734} & 0.726 & 0.718 & \underline{0.734} & 0.726 & 0.751 \\

& NDCG@3 & \textbf{0.748}${^*}$ & 0.729 & 0.728 & \underline{0.731} & 0.779 
        & \textbf{0.743} & 0.730 & 0.717 & \underline{0.733} & 0.728 & 0.717 & 0.731 & 0.727 & 0.754 \\

& NDCG@5 & \textbf{0.761}${^*}$ & 0.742 & 0.744 & \underline{0.749} & 0.785 
        & \textbf{0.755} & 0.747 & 0.735 & \underline{0.748} & 0.745 & 0.733 & \underline{0.748} & 0.745 & 0.769 \\

\midrule

\multirow{7}[0]{*}{\makecell{MSLR-\\WEB10k}}
& ERR@1 & \textbf{0.206}${^*}$ & \underline{0.161} & 0.160 & 0.148 & 0.232 
        & \textbf{0.190} & 0.134 & 0.144 & \underline{0.184} & 0.128 & 0.159 & \underline{0.184} & 0.132 & 0.203 \\

& ERR@3 & \textbf{0.282}${^*}$ & 0.230 & \underline{0.239} & 0.222 & 0.303 
        & \textbf{0.263} & 0.213 & 0.210 & \underline{0.258} & 0.204 & 0.232 & 0.257 & 0.208 & 0.278 \\

& ERR@5 & \textbf{0.305}${^*}$ & 0.255 & \underline{0.263} & 0.247 & 0.323 
        & \textbf{0.287} & 0.238 & 0.233 & \underline{0.280} & 0.229 & 0.256 & \underline{0.280} & 0.233 & 0.301 \\
\cmidrule{2-16}

& NDCG@1 & \textbf{0.493}${^*}$ & 0.421 & \underline{0.440} & 0.434 & 0.499 
        & \textbf{0.454} & 0.414 & 0.405 & \underline{0.452} & 0.398 & 0.386 & 0.450 & 0.408 & 0.465 \\

& NDCG@3 & \textbf{0.484}${^*}$ & 0.418 & \underline{0.447} & 0.437 & 0.490 
        & \textbf{0.451} & 0.430 & 0.409 & 0.446 & 0.405 & 0.397 & \underline{0.447} & 0.424 & 0.460 \\

& NDCG@5 & \textbf{0.481}${^*}$ & 0.422 & \underline{0.451} & 0.442 & 0.495 
        & \textbf{0.456} & 0.436 & 0.413 & 0.451 & 0.421 & 0.406 & \underline{0.451} & 0.431 & 0.463 \\

\midrule

\multirow{7}{*}{Istella-S}
& ERR@1 & \textbf{0.579}${^*}$ & 0.540 & 0.541 & \underline{0.549} & 0.606 
        & \textbf{0.544} & 0.523 & 0.460 & 0.523 & 0.533 & \underline{0.535} & 0.521 & 0.521 & 0.557 \\

& ERR@3 & \textbf{0.693}${^*}$ & 0.654 & 0.664 & \underline{0.668} & 0.717 
        & \textbf{0.661} & 0.645 & 0.578 & 0.645 & 0.649 & \underline{0.650} & 0.642 & 0.642 & 0.675 \\

& ERR@5 & \textbf{0.711}${^*}$ & 0.673 & 0.682 & \underline{0.688} & 0.731 
        & \textbf{0.681} & 0.665 & 0.613 & 0.666 & 0.669 & \underline{0.670} & 0.664 & 0.664 & 0.696 \\
\cmidrule{2-16}

& NDCG@1 & \textbf{0.755}${^*}$ & 0.714 & 0.727 & \underline{0.733} & 0.770 
        & \textbf{0.722} & 0.704 & 0.640 & \underline{0.706} & \underline{0.706} & \underline{0.706} & 0.701 & 0.701 & 0.733 \\

& NDCG@3 & \textbf{0.727}${^*}$ & 0.674 & 0.700 & \underline{0.705} & 0.740 
        & \textbf{0.686} & 0.673 & 0.620 & \underline{0.675} & 0.671 & 0.674 & 0.669 & 0.669 & 0.705 \\

& NDCG@5 & \textbf{0.733}${^*}$ & 0.672 & 0.705 & \underline{0.711} & 0.746 
        & \textbf{0.691} & 0.675 & 0.626 & \underline{0.679} & 0.671 & 0.677 & 0.673 & 0.673 & 0.711 \\

  \bottomrule
  \end{tabular}
  \label{tab:main_results_with_true_validation_data_short_version_detailed}
\end{table*}

\subsection{More implementation details} \label{more_implementation_details}
For the LambdaMART ranker, we choose $NDCG10$ as the optimization metric while setting the minimum number of data points in a leaf node to $2$, and the maximum number of tree leaves to $255$. The number of boosted trees and learning rate $\alpha$ are tuned in $\{{100, 300, 500\}}$ and $\{{0.05, 0.1\}}$, respectively, while all other parameters remain at their defaults. The listwise DNN model is trained for $10$k steps with a batch size of $256$ using the AdaGrad optimizer~\cite{duchi-jmlr11}. The network consists of three hidden layers of sizes [$512$, $256$, $128$] with ELU activation function~\cite{clevert-iclr16}, and is optimized using a listwise softmax cross-entropy loss. Additionally, we have a hyperparameter for the transformations of the estimated residuals used in the ranker. We tune this hyperparameter on validation data to select the best ranking model, whether using LambdaMART or DNN as the ranker. 

In the first stage (equation~\eqref{eq:control_function}), we explore Ridge, Logistic Regression, SVM, and XGBoost models to train $m(\cdot)$. Although all models yield comparable performance in the second-stage click model when their residuals are incorporated, Ridge generally performs better across all the three benchmark datasets. Therefore, we use Ridge model for learning $m(\cdot)$ for these three benchmark datasets. For the Baidu-ULTR dataset, the XGBoost model performs better in estimating residuals from $m(\cdot)$ for use in the click equation.  The impact of different choices of $m(\cdot)$ is discussed in our ablation studies. For the nonparametric hazard-ratio transformation, we estimate the PDF using fastKDE~\cite{o-csda16}. 
Unless stated otherwise, for all other experiments, we use the true relevance labels from the validation set to tune hyperparameters for both our method and the baselines. For RQ4, we use debiased clicks for hyperparameter tuning by learning $D(\cdot)$ with Ridge Regression, assuming that true relevance labels are unavailable for validation. We train $D(\cdot)$ using training data and predict on validation data. Next, we calculate the debiased clicks ($\hat{\epsilon}^{y}_{q,D}$) for validation data using the predictions from $D(\cdot)$.

\subsection{Baselines} \label{baselines}
We compare our CFC method with baseline methods from different methodological families commonly used for correcting position bias in LTR systems.
\begin{itemize}[leftmargin=9pt, nosep]
    \item $\textbf{UPE:}$ The Unconfounded Propensity Estimation~\cite{luo-sigir24} method corrects position bias with a causal method using item relevance as a confounder and addresses propensity overestimation.
    \item $\textbf{MULTR:}$ The Model-based Unbiased Learning to Rank~\cite{luo-wsdm23} method uses a context-aware simulator to generate pseudo clicks and addresses the discrepancy between pseudo and actual clicks through IPW with pseudo labels in a doubly robust manner.
    \item $\textbf{PIJD:}$ The Propensity-Independent Joint Debiasing~\cite{ovaisi-sigir21} corrects for both position and selection bias using Heckman's method.
    \item $\textbf{PairD:}$ The Pairwise Debiasing~\cite{hu-www19} model extends the IPW that jointly estimates position bias and trains a pairwise ranker.
    \item $\textbf{PAL:}$ The Position-bias Aware Learning~\cite{guo-recsys19} framework uses a two-tower model to disentangle relevance and bias factors.
    \item $\textbf{REM:}$ The regression-based EM~\cite{wang-wsdm18} model uses a position bias click model to estimate propensity scores and
    trains a ranker.
    \item $\textbf{DLA:}$ The Dual Learning Algorithm~\cite{ai-sigir18} jointly learns an unbiased propensity model and unbiased ranker from biased click data.
    \item $\textbf{LambdaMART:}$ Using LambdaMART~\cite{burges-learning10} as the LTR algorithm, we train a model directly on biased click data without any position bias correction. This serves as a baseline benchmark to assess whether our CFC method, which also uses the LambdaMART LTR algorithm as its ranker, can outperform a LambdaMART algorithm that does not correct for position bias.
    \item $\textbf{DNN:}$ We train a Deep Neural Network model directly on biased click data, following the implementation in~\cite{ai-sigir18}, without applying any position bias correction. This baseline allows us to evaluate whether our CFC method, which also uses a DNN ranker, can surpass this method that does not account for position bias.
    \item $\textbf{Oracle:}$ This model is trained using human-annotated labels. Because it relies on the true relevance of each query-item pair, its performance represents the upper bound of the ranker. Note that, while sufficiently randomizing the data may also serve as an upper bound, the Oracle method using true relevance should be the best possible upper bound. Therefore, we use this as the upper bound.
\end{itemize}

\begin{table*}
  \fontsize{7.8}{7}\selectfont
  \captionsetup{justification=raggedright, width=1.0\textwidth}
  \caption{A comparison of the overall performance of CFC and other baselines with real-world Baidu-ULTR dataset. The best-performing method for each ranker is shown in bold and the second-best is underlined.}
  \vspace{-1em}
  \centering
  \begin{tabular}{ccccccccccccccc}
    \toprule
    \multirow{2}{*}{Metrics} & \multicolumn{4}{c}{LambdaMART} & \multicolumn{7}{c}{DNN} \\ 
    \cmidrule(r){2-5} \cmidrule(r){6-13}
    & CFC & PIJD & PairD & LambdaMART 
    & CFC & PIJD & UPE & MULTR & PAL & REM & DLA & DNN \\
    \midrule

    ERR@1  & \textbf{0.140} & 0.110 & \underline{0.112} & 0.111 
           & \textbf{0.151} & 0.116 & 0.102 & 0.103 & 0.126 & \underline{0.141} & 0.091 & 0.104 \\

    ERR@3  & \textbf{0.224} & 0.183 & \underline{0.189} & 0.186 
           & \textbf{0.238} & 0.195 & 0.171 & 0.171 & 0.208 & \underline{0.226} & 0.159 & 0.172 \\

    ERR@5  & \textbf{0.251} & 0.210 & 0.211 & \underline{0.214}
           & \textbf{0.267} & 0.222 & 0.198 & 0.197 & 0.237 & \underline{0.255} & 0.187 & 0.199 \\

    \cmidrule(r){1-13}

    NDCG@1  & \textbf{0.479} & 0.381 & 0.386 & \underline{0.389} 
            & \textbf{0.511} & 0.399 & 0.375 & 0.387 & 0.425 & \underline{0.491} & 0.330 & 0.352 \\

    NDCG@3  & \textbf{0.490} & 0.402 & 0.411 & \underline{0.415} 
            & \textbf{0.521} & 0.425 & 0.397 & 0.403 & 0.453 & \underline{0.503} & 0.368 & 0.372 \\

    NDCG@5  & \textbf{0.501} & 0.417 & 0.422 & \underline{0.429} 
            & \textbf{0.530} & 0.442 & 0.408 & 0.414 & 0.472 & \underline{0.514} & 0.392 & 0.388 \\

    \bottomrule
  \end{tabular}
  \label{tab:baidu_dataset_results_detailed}
\end{table*}

\section{Additional Experimental Results} 
\subsection{Extended results of research questions} \label{sec:detailed_exp_research_qstn}
\label{more_experimental_results}
\textbf{More detailed results for RQ1}. For question RQ1, we 
compare CFC with other position bias correction methods. In Table~\ref{tab:main_results_with_true_validation_data_short_version}, we show the results based on $ERR@10$ and $NDCG@10$. Detailed results for $p = 1$, $3$, and $5$ are provided here in Table~\ref{tab:main_results_with_true_validation_data_short_version_detailed}. CFC consistently outperforms all baselines across all three datasets and all values of \( p \), whether using LambdaMART or DNN 
ranker. This shows the effectiveness of our CFC method across different metrics when evaluated at different ranking positions.

\textbf{More detailed results for RQ3}. For RQ3, we compare CFC with other position bias correction methods on the real-world Baidu industrial search engine dataset. Table~\ref{tab:baidu_dataset_results} reports results in terms of $ERR@10$ and $NDCG@10$, while detailed results for $p = 1$, $3$, and $5$ are provided in Table~\ref{tab:baidu_dataset_results_detailed}. CFC consistently outperforms all baselines across all values of $p$, under both LambdaMART and DNN rankers. This demonstrates the applicability of CFC  in a real-world large-scale industrial setting.

\begin{figure} [b]
  \centering
  \captionsetup{justification=raggedright, margin=0cm}
  \captionsetup[subfigure]{labelfont=normalfont, textfont=normalfont}
  
  \begin{subfigure}{0.48\textwidth}
    \includegraphics[width=\linewidth]{figures/legends-bold-lm.png}
  \end{subfigure}
  \begin{subfigure}{0.15 \textwidth}
    \includegraphics[width=\linewidth]{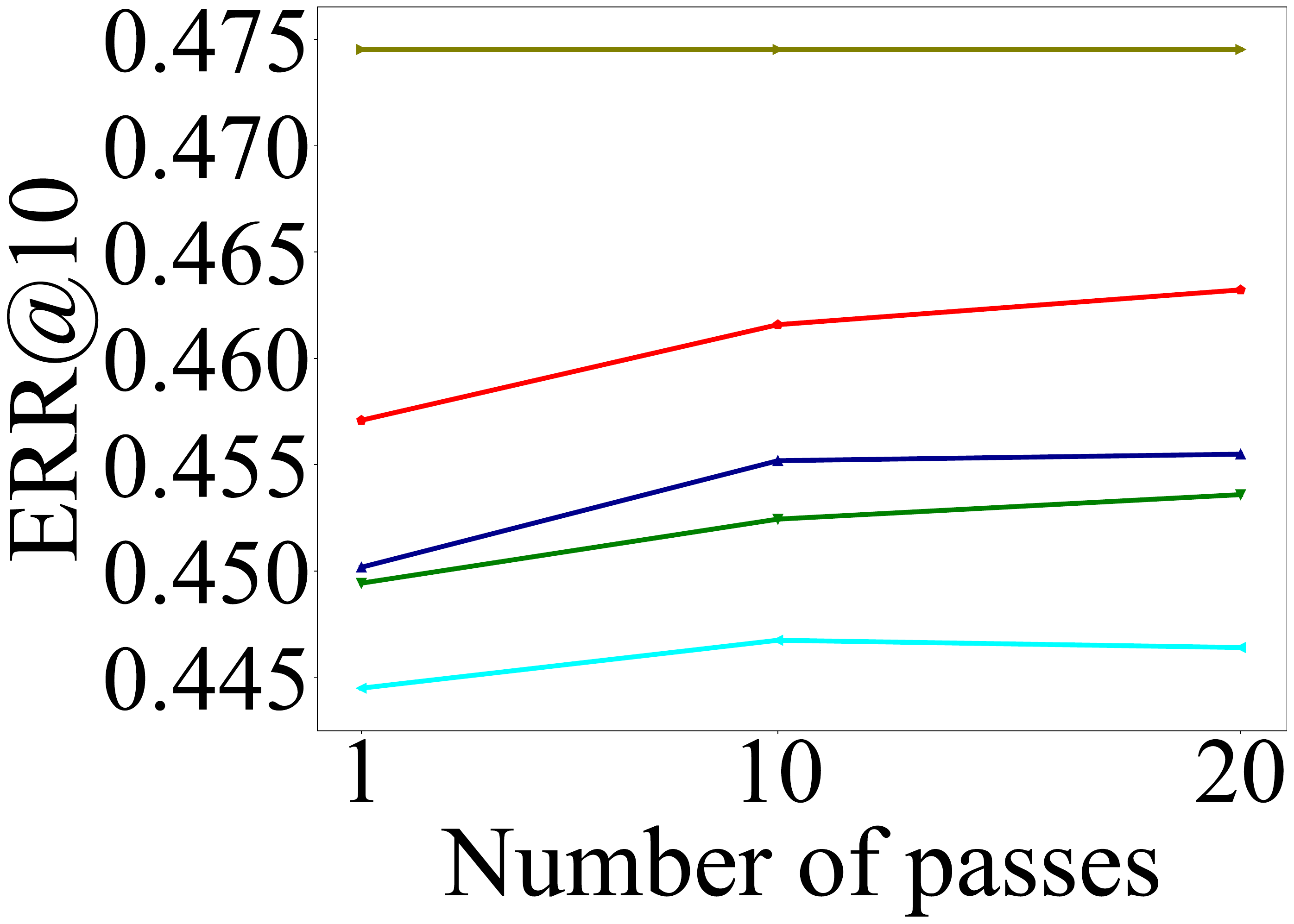}
    \caption{Yahoo}
  \end{subfigure}
  \begin{subfigure}{0.15 \textwidth}
    \includegraphics[width=\linewidth]{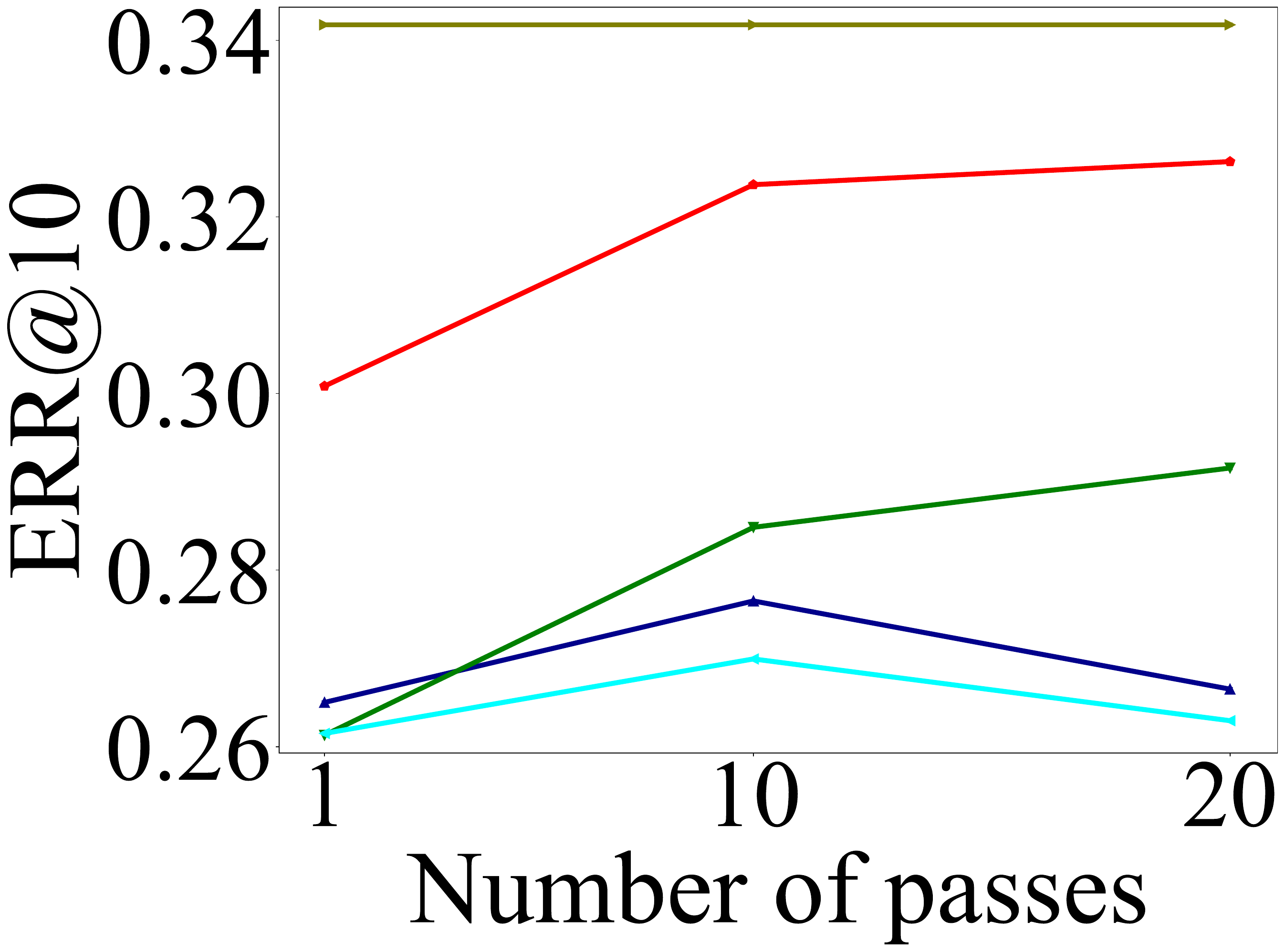}
    \caption{MSLR-WEB10k}
  \end{subfigure}
  \begin{subfigure}{0.15 \textwidth}
    \includegraphics[width=\linewidth]{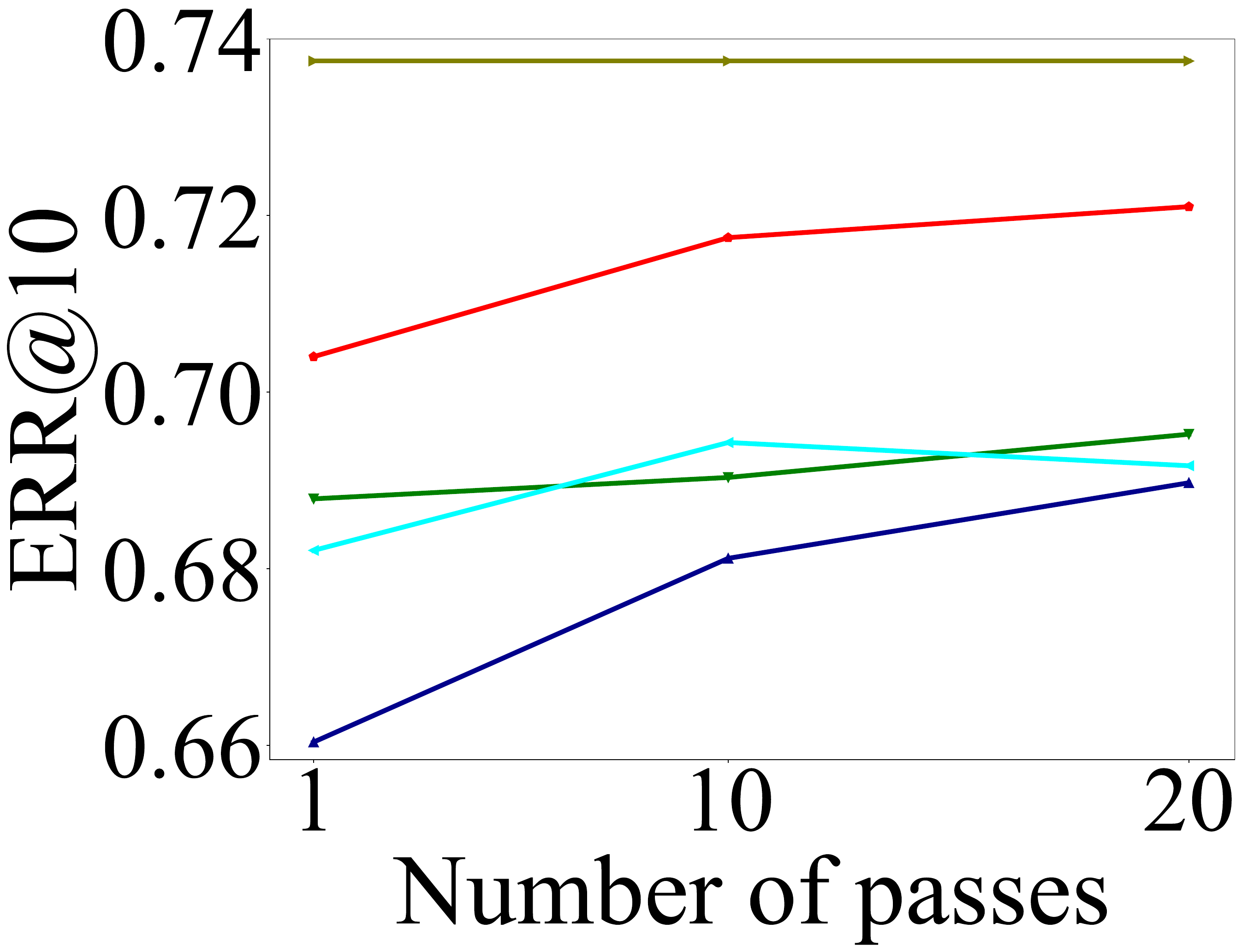}
    \caption{Istella-S}
  \end{subfigure}

  \begin{subfigure}{0.48\textwidth}
    \includegraphics[width=\linewidth]{figures/legends-bold-dnn.png}
  \end{subfigure}
  \begin{subfigure}{0.15 \textwidth}
    \includegraphics[width=\linewidth]{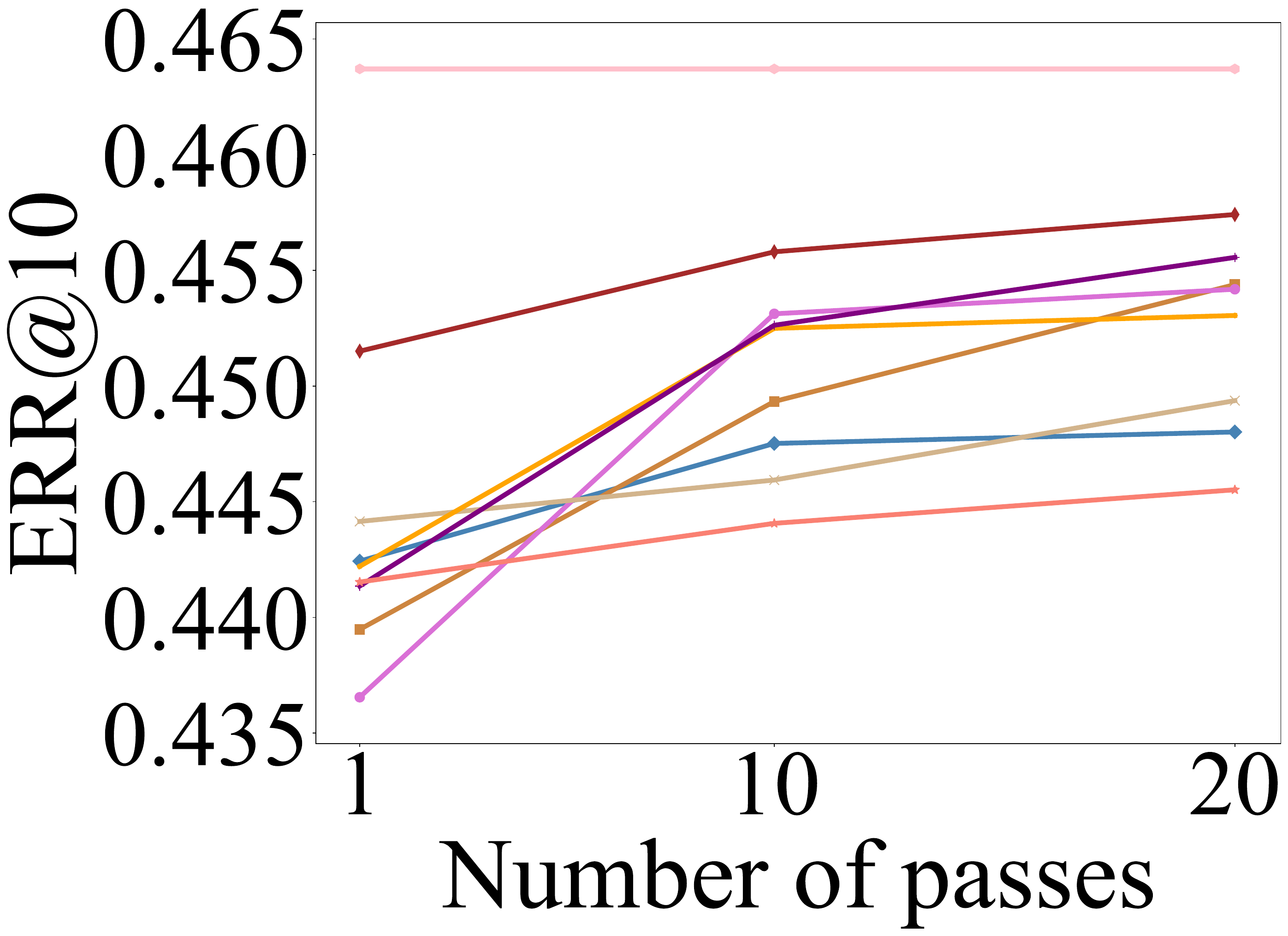}
    \caption{Yahoo}
  \end{subfigure}
  \begin{subfigure}{0.15 \textwidth}
    \includegraphics[width=\linewidth]{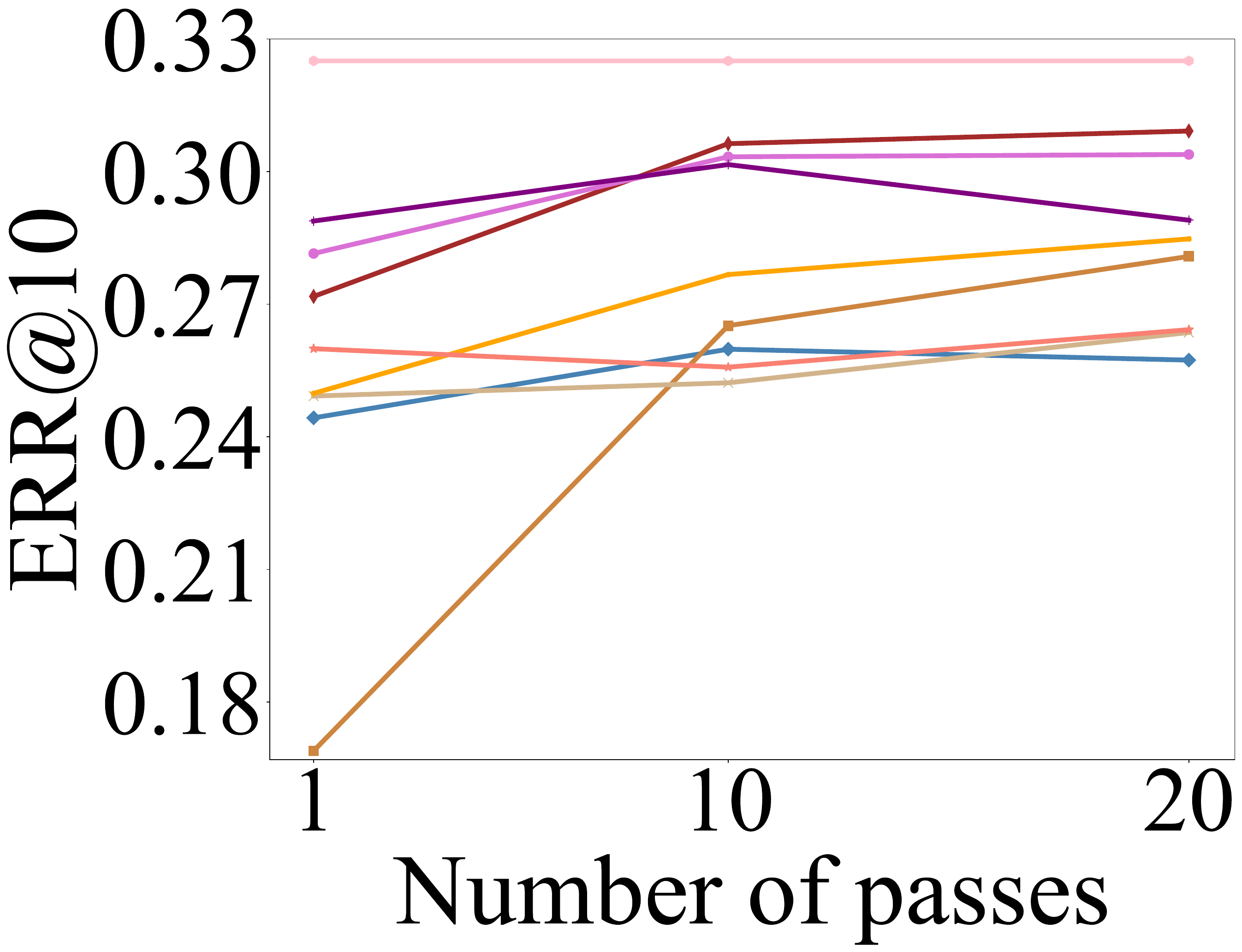}
    \caption{MSLR-WEB10k}
  \end{subfigure}
  \begin{subfigure}{0.15 \textwidth}
    \includegraphics[width=\linewidth]{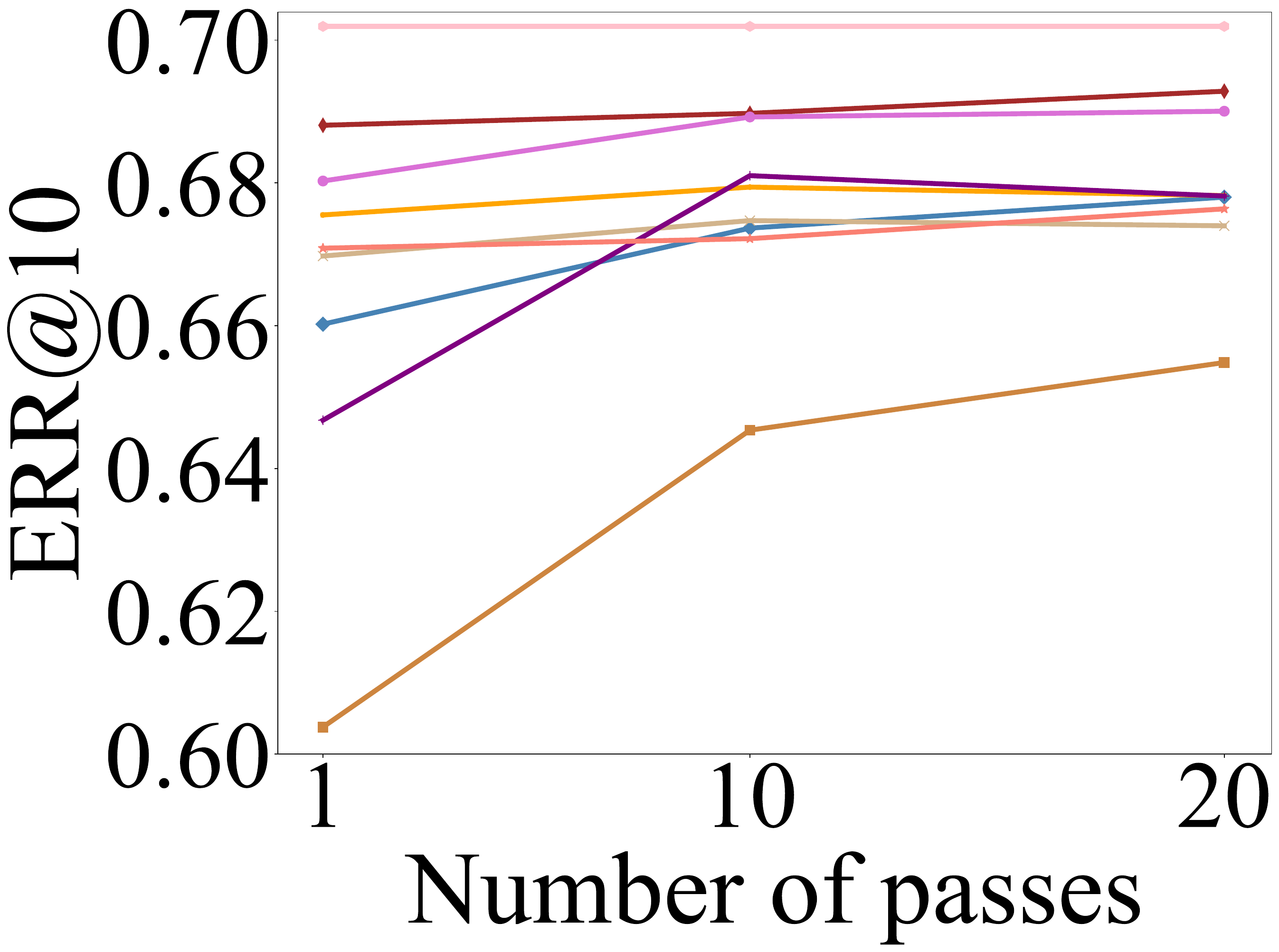}
    \caption{Istella-S}
  \end{subfigure}
  \vspace{-0.25cm}

  \captionsetup{width=0.48\textwidth}
  \caption{The performance of different methods with varying the number of passes for clicks generation.\label{fig:passes_variations-all}}
\end{figure}

\begin{table*} [hb]
  \setlength{\tabcolsep}{4.8pt} 
  \fontsize{7.8}{7}\selectfont
  \captionsetup{justification=raggedright, width=1.0\textwidth}
    \caption{A comparison of the overall performance of CFC and other baselines under stochastic logging policy. The best-performing method for each ranker, excluding Oracle, is shown in bold, and the second-best is underlined. * denotes a statistically significant improvement of the best-performing method over the best baseline, considering all LambdaMART and DNN rankers, with ${\bm{p < 0.05}}$.}
  \vspace{-1em}
  \centering
  \scalebox{1.0}{%
    \begin{tabular}{cccccccccccccccc}
    \toprule
    \multirow{2}{*}{Datasets} & \multirow{2}{*}{Metrics} & \multicolumn{5}{c} {LambdaMART} & \multicolumn{9}{c}{DNN} \\ \cmidrule(r){3-7} \cmidrule(r){8-16}
        & & CFC & PIJD & PairD & LambdaMART & Oracle & CFC & PIJD & UPE & MULTR & PAL & REM & DLA & DNN & Oracle \\
     \midrule

    \multirow{9}[0]{*}{Yahoo}
    & ERR@1 & \textbf{0.348}${^*}$ & \underline{0.342} & 0.337 & 0.330 & 0.359 & \textbf{0.342} & 0.336 & 0.337 & \underline{0.341} & 0.331 & 0.338 & 0.337 & 0.330 & 0.346 \\
    & ERR@3 & \textbf{0.427}${^*}$ & \underline{0.417} & 0.417 & 0.411 & 0.438 & \textbf{0.422} & 0.416 & 0.414 & \underline{0.418} & 0.411 & 0.414 & 0.416 & 0.411 & 0.426 \\
    & ERR@5 & \textbf{0.448}${^*}$ & \underline{0.438} & 0.437 & 0.434 & 0.460 & \textbf{0.444} & 0.439 & 0.435 & \underline{0.440} & 0.434 & 0.436 & 0.438 & 0.433 & 0.449 \\
    & ERR@10 & \textbf{0.464}${^*}$ & \underline{0.455} & 0.453 & 0.449 & 0.474 & \textbf{0.459} & 0.454 & 0.452 & \underline{0.455} & 0.450 & 0.452 & 0.453 & 0.448 & 0.464 \\
    \cmidrule(r){2-16}
    & NDCG@1 & \textbf{0.754}${^*}$ & 0.733 & \underline{0.737} & 0.737 & 0.778 & \textbf{0.746} & 0.734 & 0.715 & 0.737 & 0.729 & 0.717 & \underline{0.738} & 0.734 & 0.751 \\
    & NDCG@3 & \textbf{0.750}${^*}$ & 0.728 & \underline{0.729} & 0.734 & 0.779 & \textbf{0.745} & 0.734 & 0.717 & \underline{0.738} & 0.728 & 0.718 & 0.734 & 0.732 & 0.754 \\
    & NDCG@5 & \textbf{0.764}${^*}$ & 0.744 & \underline{0.747} & 0.750 & 0.785 & \textbf{0.760} & 0.750 & 0.735 & \underline{0.753} & 0.746 & 0.733 & 0.751 & 0.750 & 0.769 \\
    & NDCG@10 & \textbf{0.802}${^*}$ & 0.786 & \underline{0.791} & 0.789 & 0.817 & \textbf{0.797} & 0.788 & 0.777 & \underline{0.791} & 0.786 & 0.775 & 0.790 & 0.787 & 0.805 \\
    \midrule

    \multirow{9}[0]{*}{\makecell{MSLR-\\WEB10k}}
    & ERR@1 & \textbf{0.199}${^*}$ & 0.158 & \underline{0.163} & 0.156 & 0.232 & \textbf{0.198} & 0.139 & 0.151 & \underline{0.184} & 0.154 & 0.119 & 0.174 & 0.148 & 0.203 \\
    & ERR@3 & \textbf{0.275}${^*}$ & 0.228 & \underline{0.243} & 0.228 & 0.303 & \textbf{0.270} & 0.218 & 0.215 & \underline{0.258} & 0.225 & 0.191 & 0.250 & 0.223 & 0.278 \\
    & ERR@5 & \textbf{0.297}${^*}$ & 0.253 & \underline{0.267} & 0.253 & 0.323 & \textbf{0.295} & 0.242 & 0.240 & \underline{0.278} & 0.248 & 0.217 & 0.274 & 0.249 & 0.301 \\
    & ERR@10 & \textbf{0.317}${^*}$ & 0.275 & \underline{0.288} & 0.275 & 0.342 & \textbf{0.315} & 0.264 & 0.261 & \underline{0.300} & 0.270 & 0.240 & 0.294 & 0.269 & 0.325 \\
    \cmidrule(r){2-16}
    & NDCG@1 & \textbf{0.482}${^*}$ & 0.421 & 0.443 & \underline{0.448} & 0.499 & \textbf{0.471} & 0.421 & 0.382 & 0.441 & 0.369 & 0.363 & \underline{0.456} & 0.430 & 0.465 \\
    & NDCG@3 & \textbf{0.476}${^*}$ & 0.419 & \underline{0.450} & 0.446 & 0.490 & \textbf{0.460} & 0.440 & 0.379 & \underline{0.445} & 0.374 & 0.379 & 0.453 & 0.435 & 0.460 \\
    & NDCG@5 & \textbf{0.475}${^*}$ & 0.423 & \underline{0.455} & 0.449 & 0.495 & \textbf{0.467} & 0.442 & 0.387 & 0.446 & 0.385 & 0.388 & \underline{0.457} & 0.441 & 0.463 \\
    & NDCG@10 & \textbf{0.487}${^*}$ & 0.439 & \underline{0.468} & 0.463 & 0.498 & \textbf{0.477} & 0.454 & 0.403 & 0.461 & 0.407 & 0.411 & \underline{0.467} & 0.454 & 0.476 \\
    \midrule

    \multirow{9}[0]{*}{Istella-S}
    & ERR@1 & \textbf{0.575}${^*}$ & 0.518 & 0.542 & \underline{0.556} & 0.606 & \textbf{0.566} & 0.516 & 0.463 & 0.560 & 0.533 & \underline{0.644} & 0.543 & 0.535 & 0.557 \\
    & ERR@3 & \textbf{0.690}${^*}$ & 0.641 & 0.667 & \underline{0.675} & 0.717 & \textbf{0.678} & 0.641 & 0.581 & \underline{0.673} & 0.653 & 0.655 & 0.655 & 0.653 & 0.675 \\
    & ERR@5 & \textbf{0.708}${^*}$ & 0.660 & 0.687 & \underline{0.695} & 0.731 & \textbf{0.696} & 0.662 & 0.615 & \underline{0.691} & 0.673 & 0.674 & 0.676 & 0.673 & 0.696 \\
    & ERR@10 & \textbf{0.715}${^*}$ & 0.669 & 0.694 & \underline{0.702} & 0.737 & \textbf{0.703} & 0.670 & 0.646 & \underline{0.699} & 0.681 & 0.683 & 0.684 & 0.680 & 0.702 \\
    \cmidrule(r){2-16}
    & NDCG@1 & \textbf{0.753}${^*}$ & 0.688 & 0.730 & \underline{0.741} & 0.770 & \textbf{0.736} & 0.698 & 0.644 & \underline{0.726} & 0.714 & 0.708 & 0.704 & 0.714 & 0.733 \\
    & NDCG@3 & \textbf{0.723}${^*}$ & 0.659 & 0.703 & \underline{0.711} & 0.740 & \textbf{0.697} & 0.670 & 0.625 & \underline{0.691} & 0.682 & 0.664 & 0.669 & 0.679 & 0.705 \\
    & NDCG@5 & \textbf{0.729}${^*}$ & 0.663 & 0.710 & \underline{0.717} & 0.746 & \textbf{0.699} & 0.676 & 0.631 & \underline{0.693} & 0.685 & 0.662 & 0.672 & 0.683 & 0.711 \\
    & NDCG@10 & \textbf{0.766}${^*}$ & 0.690 & 0.745 & \underline{0.753} & 0.777 & \textbf{0.727} & 0.706 & 0.662 & \underline{0.722} & 0.718 & 0.682 & 0.700 & 0.711 & 0.745 \\

    \bottomrule
    \end{tabular}
    }
    \label{tab:main_results_with_true_validation_data_short_version_stochastic_detailed}
\end{table*}

\textbf{Performance trends with increasing clicks (RQ5).} \label{sec:passes_variations} We evaluate the performance of the CFC method by varying the amount of click data. To achieve this, we generate clicks by performing 1, 10, and 20 passes over the training data. For $1$, $10$, and $20$ training passes, we obtain approximately $13$k, $127$k, and $253$k clicks on Yahoo dataset; $3$k, $27$k, and $54$k on MSLR-WEB10k dataset; and $22$k, $223$k, and $446$k on Istella-S dataset. Figure~\ref{fig:passes_variations-all} shows the $ERR@10$ performance with LambdaMART and DNN as rankers. Across all three datasets and all passes, CFC with LambdaMART consistently outperforms all baseline methods. The performance of CFC with DNN also improves with the number of clicks, achieving the best results on all the datasets among all DNN-based baselines, except for pass $1$ on MSLR-WEB10K dataset. As the number of clicks (passes) increases, the performance of our method improves and gets closer to the performance of the Oracle method. With more clicks (passes), the model benefits from a bias-variance tradeoff, with reduced bias from unbiased method and lower variance due to more stable training data, leading to improved performance.

\begin{figure}
  \centering
  \captionsetup{justification=raggedright, margin=0cm}
  \captionsetup[subfigure]{labelfont=normalfont, textfont=normalfont}

  \begin{subfigure}{0.48\textwidth}
    \includegraphics[width=\linewidth]{figures/legends-bold-lm.png}
  \end{subfigure}
  \begin{subfigure}{0.15 \textwidth}
    \includegraphics[width=\linewidth]{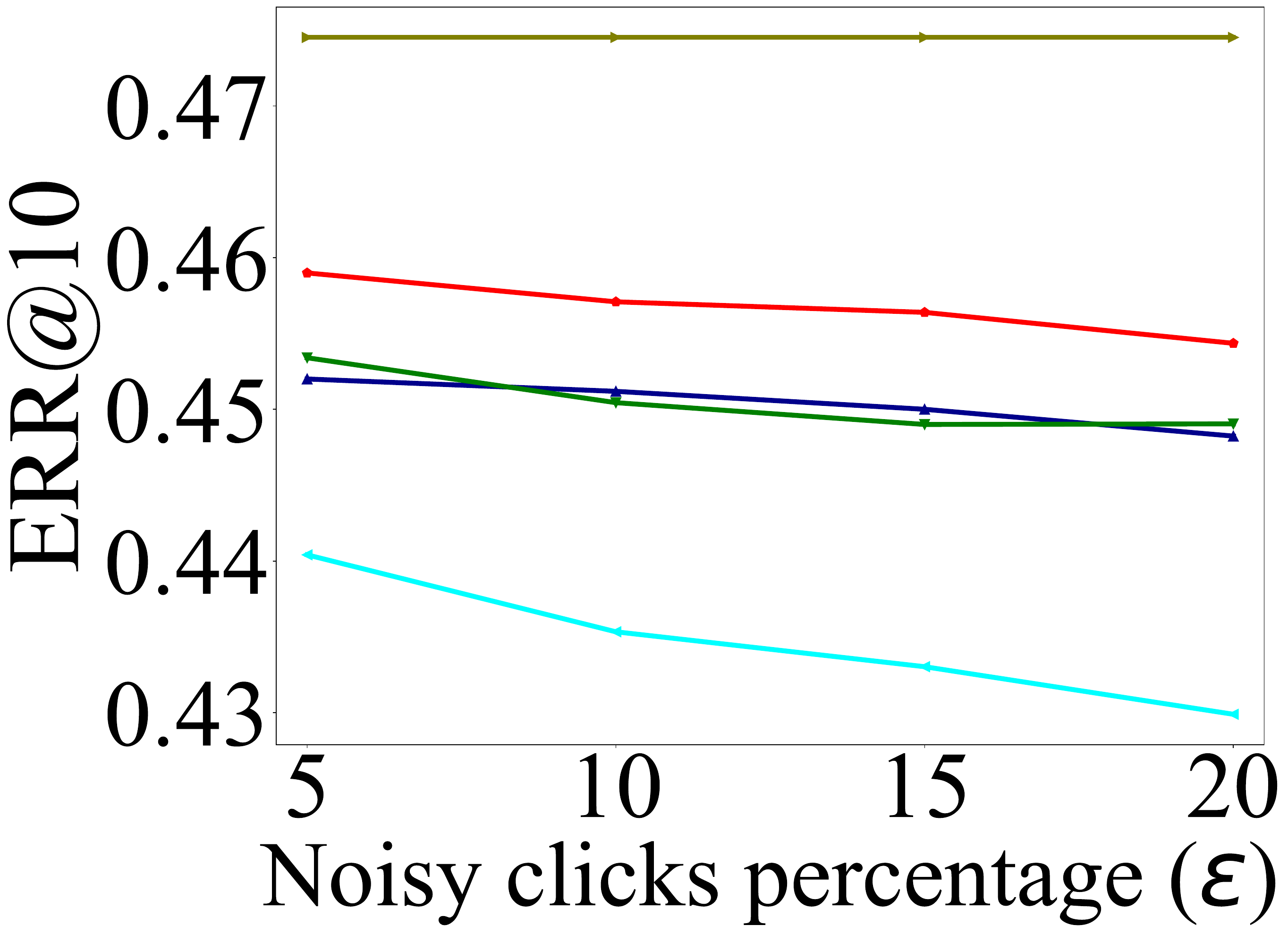}
    \caption{Yahoo}
  \end{subfigure}
  \begin{subfigure}{0.15 \textwidth}
    \includegraphics[width=\linewidth]{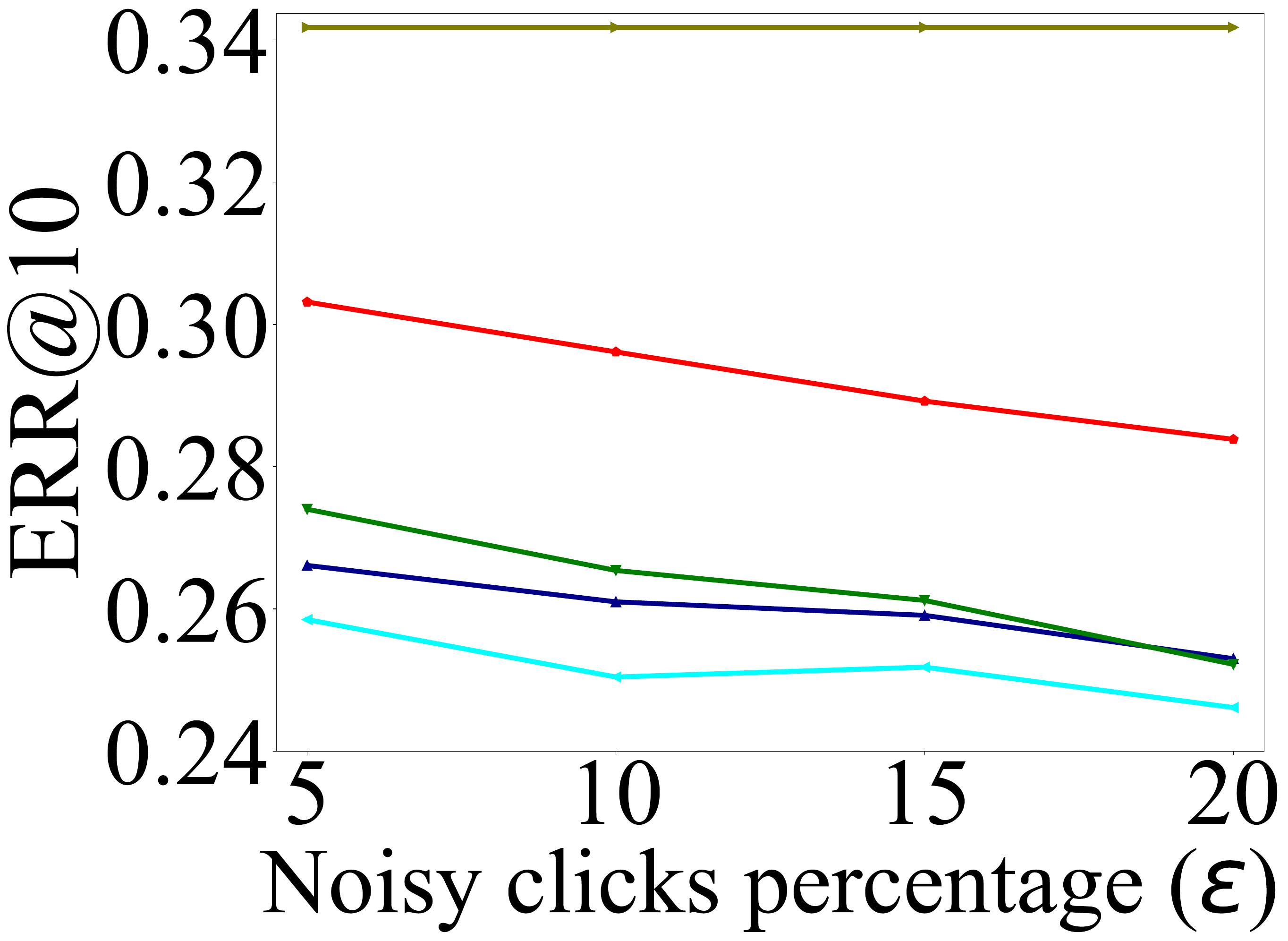}
    \caption{MSLR-WEB10k}
  \end{subfigure}
  \begin{subfigure}{0.15 \textwidth}
    \includegraphics[width=\linewidth]{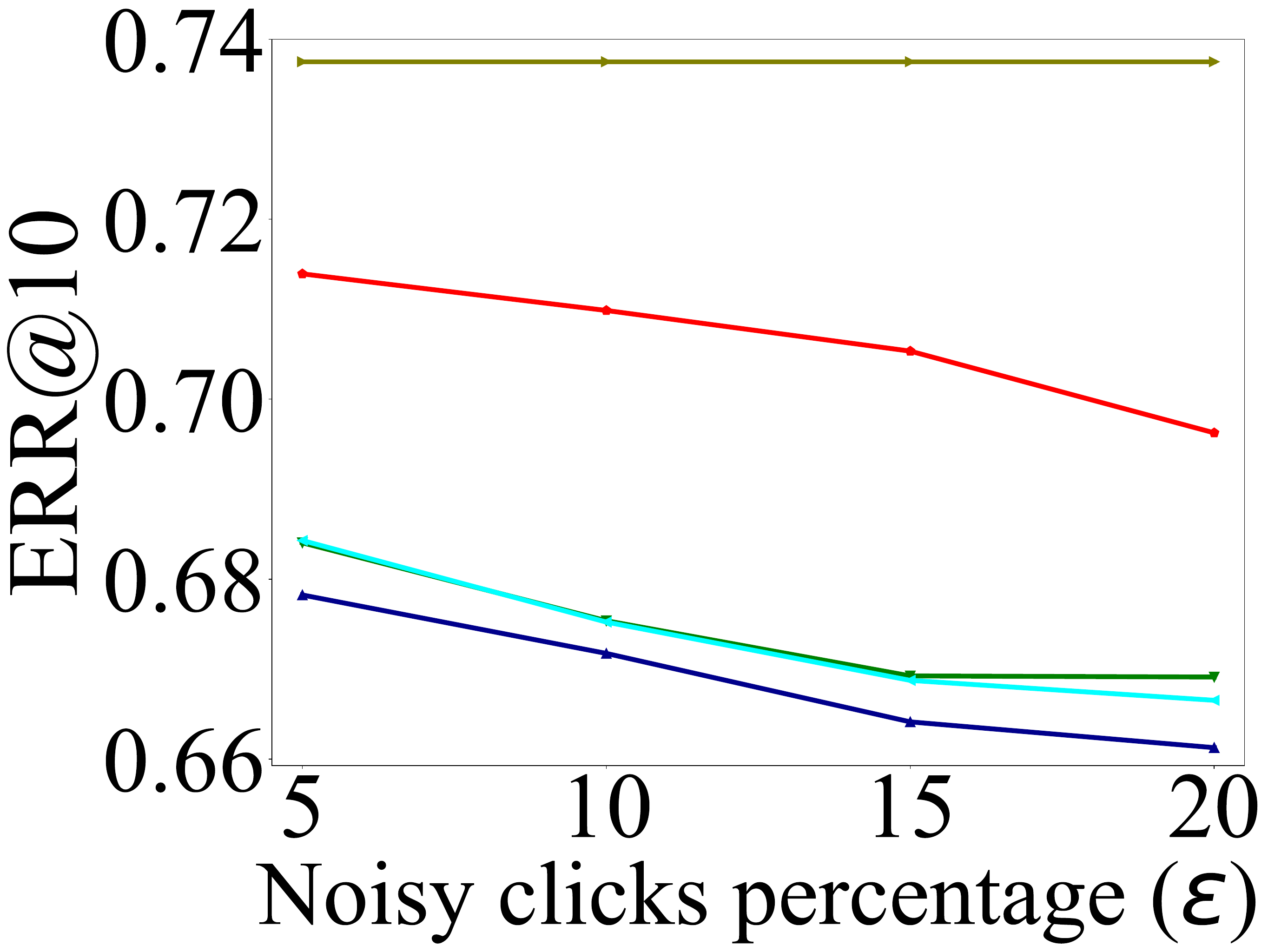}
    \caption{Istella-S}
  \end{subfigure}

  \begin{subfigure}{0.48\textwidth}
    \includegraphics[width=\linewidth]{figures/legends-bold-dnn.png}
  \end{subfigure}
  \begin{subfigure}{0.15 \textwidth}
    \includegraphics[width=\linewidth]{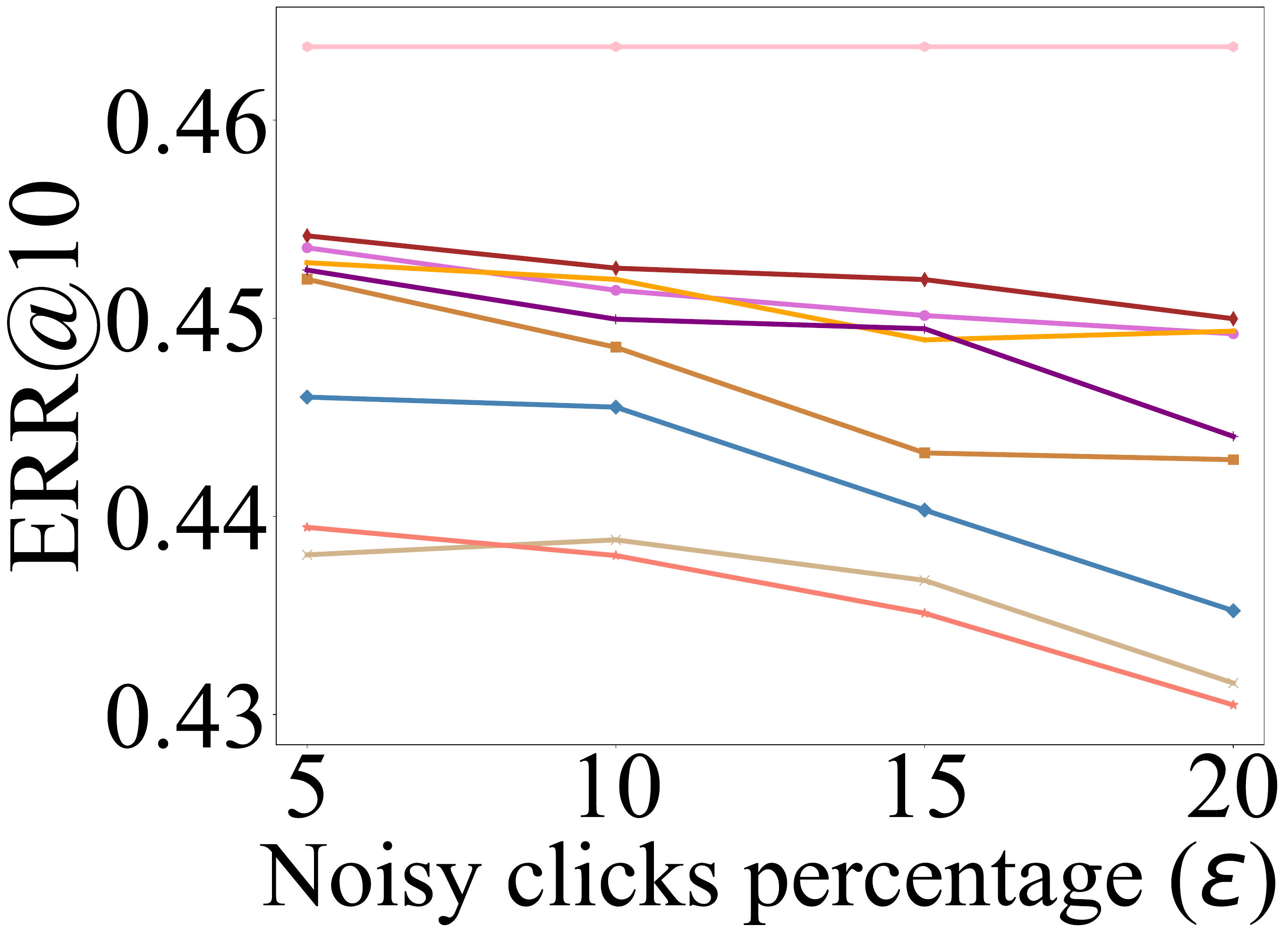}
    \caption{Yahoo}
  \end{subfigure}
  \begin{subfigure}{0.15 \textwidth}
    \includegraphics[width=\linewidth]{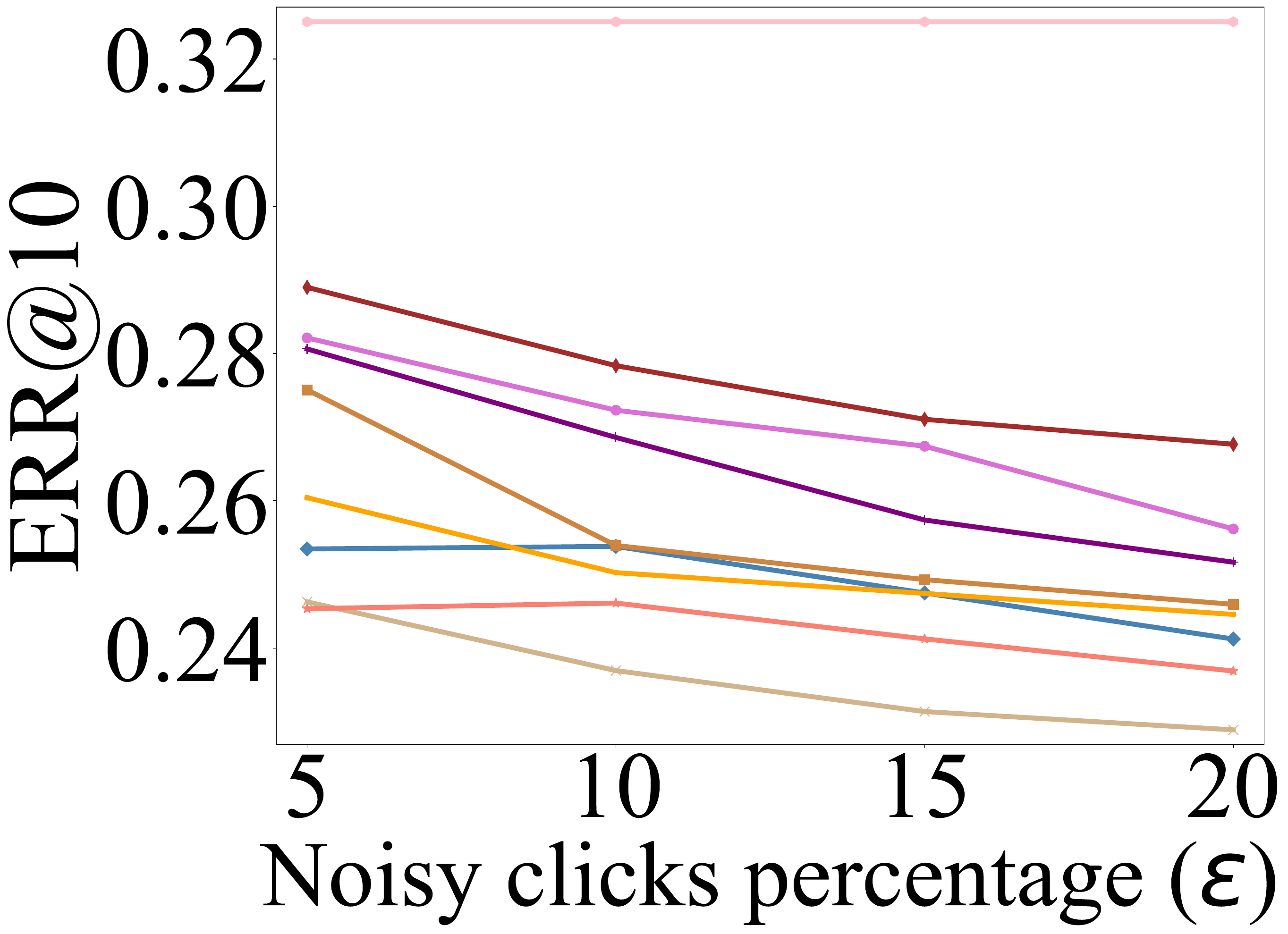}
    \caption{MSLR-WEB10k}
  \end{subfigure}
  \begin{subfigure}{0.15 \textwidth}
    \includegraphics[width=\linewidth]{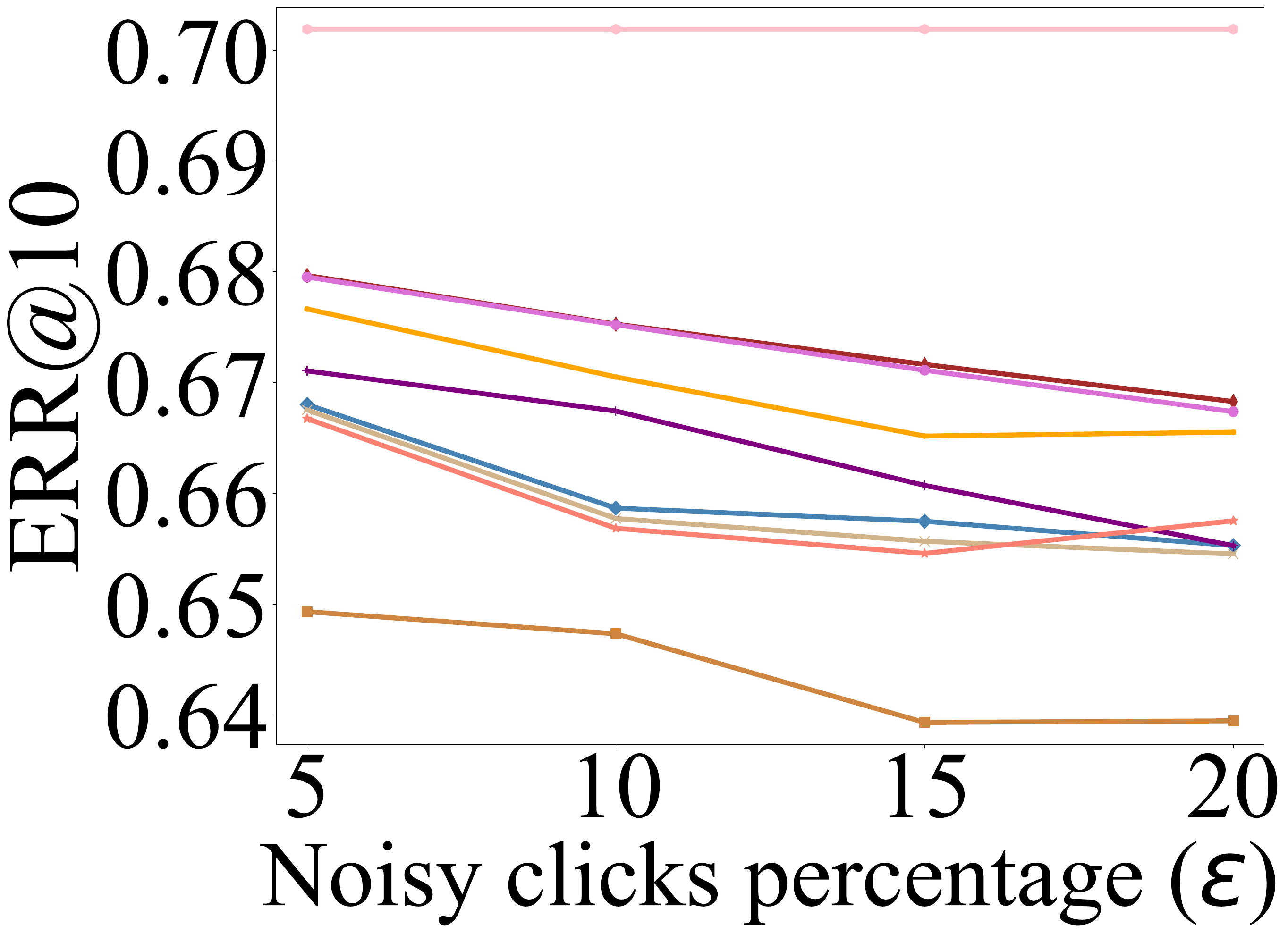}
    \caption{Istella-S}
  \end{subfigure}
  \vspace{-0.25cm}

  \captionsetup{width=0.48\textwidth}
  \caption{The performance of different methods up to $\bm{10 \ (p)}$ positions with varying percentages of noisy clicks.
  \label{fig:noisy_clicks}}
\end{figure}

\textbf{The effect of click noise (RQ6).}
Noisy clicks commonly occur in real-world settings when users interact with irrelevant items. To assess robustness to such noise, we vary the probability of clicking on irrelevant items ($\epsilon = 5\%, 10\%, 15\%, 20\%$). The $ERR@10$ results for both LambdaMART and DNN rankers are shown in Figure~\ref{fig:noisy_clicks}.
Across all three datasets and under all noise levels, CFC with LambdaMART consistently outperforms all baseline methods. CFC also achieves the best performance on every dataset among all DNN-based baselines, except that it approximately matches MULTR on Istella-S at noise levels of 5\% and 10\%. Therefore, our method demonstrates greater robustness to noisy clicks compared to baselines using either LambdaMART or DNN as the ranker.

\textbf{Performance of CFC and other methods under stochastic logging policy (RQ7).} In response to RQ7, we evaluate CFC against baselines in simulation experiments under stochastic logging policy. We report the results in Table~\ref{tab:main_results_with_true_validation_data_short_version_stochastic_detailed}. Across all three benchmark datasets, CFC consistently outperforms the corresponding baseline methods in terms of both ERR and NDCG for LambdaMART and DNN rankers across different values of $p$. These findings highlight the effectiveness of CFC under stochastic logging settings. Under the stochastic logging policy, CFC generally achieves better performance, particularly with the DNN ranker, compared to the deterministic logging policy shown in Table~\ref{tab:main_results_with_true_validation_data_short_version} and Table~\ref{tab:main_results_with_true_validation_data_short_version_detailed} for RQ1. This occurs because the stochastic logging policy introduces variation in item rankings, leading to greater variation in the first-stage residuals, which in turn enables more accurate estimation of feature parameters and improves the ranking performance of CFC method. Note that under the stochastic logging policy, the performance of baseline methods also improves slightly in most cases compared to the deterministic setting, particularly for IPW-based methods such as MULTR, UPE, and DNN. This improvement is due to the introduced randomization, which enables more accurate propensity estimation and, in turn, enhances ranking performance. However, CFC still consistently outperforms these baselines under stochastic logging settings.

\subsection{Detailed ablation studies} \label{more_ablations}
\textbf{Robustness to production ranker variations.}
In our experiments, we use 1\% of the data with true relevance labels to train the production ranker (RankSVM) that generates the ranked list $\pi_q$. Now, we vary the production ranker by changing the amount of true relevance data used for its training. We use 10\%, 20\%, 30\%, and 100\% of the train data to train it. We report the results using LambdaMART as the ranker in Table~\ref{tab:prod_variations}. 
The results show that the performance remains similar across different production rankers for all datasets, highlighting the robustness of our method to the specification of production ranker used to generate the ranked lists for position-bias correction. Moreover, as the amount of training data for the production ranker increases, it becomes a stronger ranker (i.e., it produces higher-quality rankings). Even under a stronger logging policy (e.g., 100\% data), the CFC method achieves performance comparable to that obtained with a weaker production ranker (e.g., trained with only 1\% of the data). This trend is also observed in our experiments on the Baidu dataset, where the production ranker is strong, yet our method continues to perform well.
\begin{table} [hb]
  \fontsize{7.8}{7}\selectfont
  \captionsetup{justification=raggedright, width=0.49\textwidth}
  \caption{The performance of CFC using LambdaMART by varying the percentage of labeled data used for the production ranker.}
  \vspace{-1em}
  \centering
  \begin{tabular}{cccccc}
    \toprule
    \multirow{2}{*}{Datasets} & \makecell{1\%} & \makecell{10\%} & \makecell{20\%} & \makecell{30\%} & \makecell{100\%} \\ \cmidrule(r){2-6} & \multicolumn{5}{c} {$ERR@10$} \\
    \midrule
    Yahoo & \textbf{0.462} & \textbf{0.462} & 0.459 & 0.458 & \textbf{0.462} \\
    \makecell{MSLR-WEB10k} & 0.324 & 0.316 & 0.314 & 0.316 & \textbf{0.332} \\
    Istella-S & \textbf{0.717} & 0.710 & 0.716 & 0.714 & 0.712 \\
    \bottomrule
  \end{tabular}
  \label{tab:prod_variations}
\end{table}

\begin{table}
  \fontsize{7.8}{7}\selectfont
  \captionsetup{justification=raggedright, width=0.49\textwidth}
    \caption{A comparison between CFC and CFC-S across various datasets.}
  \vspace{-1em}
  \centering
    \begin{tabular}{cccccc}
    \toprule
    \multirow{2}{*}{Datasets} & \multirow{2}{*}{Metrics} & \multicolumn{2}{c} {LambdaMART} & \multicolumn{2}{c}{DNN} \\ \cmidrule(r){3-4} \cmidrule(r){5-6}
        & & CFC & CFC-S & CFC & CFC-S \\
     \midrule

    \multirow{2}[0]{*}{Yahoo}
    & ERR@10 & \textbf{0.462} & \textbf{0.462} & \textbf{0.456} & \textbf{0.456} \\
    & NDCG@10 & \textbf{0.801} & 0.796 & \textbf{0.794} & 0.792 \\
    \midrule

    \multirow{2}[0]{*}{\makecell{MSLR-\\WEB10k}}
    & ERR@10 & \textbf{0.324} & 0.317 & \textbf{0.306} & \textbf{0.306} \\
    & NDCG@10 & \textbf{0.489} & 0.486 & \textbf{0.467} & \underline{0.466} \\
    \midrule

    \multirow{2}[0]{*}{Istella-S}
    & ERR@10 & \textbf{0.717} & 0.714 & \textbf{0.689} & 0.683 \\
    & NDCG@10 & \textbf{0.766} & 0.765 & \textbf{0.721} & 0.712 \\

    \bottomrule
    \end{tabular}
    \label{tab:cfc_cfc_s_diff}
\end{table}
\textbf{Residual-only control function}
We also conduct a study using only the residuals in the second-stage click equation, instead of employing the full set of Lewbel’s interaction terms with features. Specifically, if we set the control function as $\Phi\left( \bm{x}_q^y, T(\hat{\epsilon}_{q,m}^y) \right) = T(\hat{\epsilon}_{q,m}^y)$, the click equation based on Lewbel’s formulation in equation~\eqref{eq:unbiased_c_lewbel} simplifies to
$C_q^y \;=\; \mathbb{I}\!\left[s\big(\bm{x}_q^y, T(\hat{\epsilon}^{y}_{q,m})\big) + \epsilon^y_{q,g} > 0\right].$
We refer to this simplified variant as CFC-S. This formulation directly uses the residuals in the second-stage click equation.
We report the performance of CFC and CFC-S in Table~\ref{tab:cfc_cfc_s_diff} using both LambdaMART and DNN rankers. Across all three datasets, CFC consistently outperforms CFC-S. However, on the Yahoo dataset, CFC-S performs comparably to CFC on the $ERR@10$ metric. Similarly, on the MSLR-WEB10k dataset with the DNN ranker, both methods achieve identical performance on $ERR@10$. In all other cases, CFC yields better results than CFC-S. Overall, CFC-S remains competitive and performs close to CFC in most settings. One possible explanation is that the objective of an LTR system is to improve predictive accuracy by correcting position bias and recovering well-calibrated relevance scores, rather than estimating the causal effect of rank. When the first-stage residuals adequately capture the dependence between rank and clicks, 
{conditioning solely on the ranking residual can still improve predictive performance, even though, as discussed in Section~\ref{control_function_section}, the residual-only specification does not recover unbiased feature parameters without the heteroskedasticity-based terms.}

{Because the control-function terms are set to zero at inference for CFC-S, any predictive gains are coming from the estimated feature effects. Conditioning on the ranking residual absorbs confounding due to variation in the production ranker's decisions that is not explained by the observed features---e.g., personalization or other unobserved information used by the production ranker. It does not, however, remove position bias associated with the feature-predicted component of rank, $m(\bm{x}_q^y)$, because the same features also enter the click model, making feature-driven relevance (for clicks) difficult to separate from  the position assigned to those features. Separating these effects would require exogenous variation in position, such as the randomized exposure considered in Table~\ref{tab:main_results_with_true_validation_data_short_version_stochastic_detailed}. When the ranking residual is heteroskedastic, the full CFC specification adds residual--feature interactions that help capture how unexplained ranking variation changes across the feature space, which can explain its additional gains over CFC-S.}

\begin{table} [ht]
  \fontsize{7.8}{7}\selectfont
  \captionsetup{justification=raggedright, width=0.49\textwidth}
  \caption{The performance of the CFC method using LambdaMART with different machine learning algorithms used to estimate the ranking process in first stage.}
  \vspace{-1em}
  \centering
  \begin{tabular}{ccccc}
    \toprule
    \multirow{2}{*}{Datasets} & \makecell{Ridge} & \makecell{Logistic Regression} & \makecell{SVM} & \makecell{XGBoost} \\ \cmidrule(r){2-5} & \multicolumn{4}{c} {$ERR@10$} \\
    \midrule
    Yahoo & \textbf{0.462} & 0.460 & 0.458 & 0.456 \\
    \makecell{MSLR-WEB10k} & \textbf{0.324} & 0.312 & 0.307 & 0.308 \\
    Istella-S & \textbf{0.717} & 0.713 & 0.716 & 0.710 \\
    \bottomrule
  \end{tabular}
  \label{tab:cf_variations}
\end{table}
\textbf{The effect of model specification in estimating the ranking process.} We investigate how the specification of machine learning model used to train $m(\cdot)$—from which the residuals are estimated and used in the second stage—affects the ranking performance of the CFC method. We experiment with four machine learning models to train $m(\cdot)$: Ridge, Logistic Regression, SVM, and XGBoost. The performance of the CFC method with LambdaMART using each of these models for learning $m(\cdot)$ is shown in Table~\ref{tab:cf_variations}. Our results show that all models used to estimate $m(\cdot)$ yield comparable ranking performance for second stage model, indicating that the CFC method is robust and relatively insensitive to potential misspecification of the first-stage model. However, using Ridge for the control function yields the best ranking performance on all datasets. 
These findings suggest that the effectiveness of our approach is not heavily dependent on the specific choice of the first-stage model.
Moreover, regarding the computational efficiency of the first stage model, the models used (e.g., Ridge and Logistic Regression) are lightweight and highly scalable. XGBoost is also known for its efficiency and ability to handle large-scale datasets. The Ridge model, which achieves the best performance—incurs minimal computational overhead, training in 10.4 seconds on Yahoo, 7.4 seconds on MSLR-WEB10K, and 15.3 seconds on Istella-S datasets. Overall, our first stage can leverage simple but efficient model to compute residuals, resulting 
in much lower computational overhead.

\begin{table} [ht]
  \fontsize{7.8}{7}\selectfont
  \captionsetup{justification=raggedright, width=\columnwidth}
  \caption{Residual statistics of the first-stage ranking model across production rankers trained on varying percentages of labeled data. We report the mean residual, variance, and Fligner--Killeen (FK) statistic for each production ranker. FK values are statistically significant with $\bm{p < 0.05}$ across all setups.}
  \vspace{-1em}
  \centering
  \begin{tabular}{ccccc}
    \toprule
    Datasets & Setup & Mean & Var & FK \\
    \midrule
    \multirow{5}{*}{Yahoo}
      & 1\%   & 0.096 & 0.0105 & 81{,}620 \\
      & 10\%  & 0.103 & 0.0120 & 79{,}852 \\
      & 20\%  & 0.100 & 0.0130 & 90{,}660 \\
      & 30\%  & 0.085 & 0.0118 & 76{,}837 \\
      & 100\% & 0.076 & 0.0088 & 66{,}574 \\
    \midrule
    \multirow{5}{*}{MSLR-WEB10k}
      & 1\%   & 0.061 & 0.0051 & 34{,}178 \\
      & 10\%  & 0.058 & 0.0049 & 20{,}554 \\
      & 20\%  & 0.058 & 0.0052 & 22{,}861 \\
      & 30\%  & 0.067 & 0.0065 & 23{,}425 \\
      & 100\% & 0.049 & 0.0036 & 19{,}243 \\
    \midrule
    \multirow{5}{*}{Istella-S}
      & 1\%   & 0.134 & 0.0155 & 3{,}714 \\
      & 10\%  & 0.134 & 0.0157 & 4{,}068 \\
      & 20\%  & 0.142 & 0.0176 & 4{,}131 \\
      & 30\%  & 0.149 & 0.0195 & 3{,}009 \\
      & 100\% & 0.128 & 0.0162 & 2{,}082 \\
    \bottomrule
  \end{tabular}
  \label{tab:prod_residual_variations}
\end{table}

\textbf{Residual analysis across different production rankers.}
We estimate the residual statistics for the first-stage ranking model $m(\cdot)$ across different production rankers that generate the rankings and clicks, trained on varying fractions of labeled data.
Table~\ref{tab:prod_residual_variations} shows the mean, variance, and Fligner-Killeen (FK)~\cite{fligner-jasa76} statistic of the residuals from the first-stage ranking model. As the amount of training data increases, the rankers learn more effectively from the data and become stronger predictors through better estimation of model parameters. Across all three datasets, stronger production rankers produce smaller residuals: the 100\% labeled data setting yields the lowest mean residual on Yahoo (0.076), MSLR-WEB10k (0.049), and Istella-S (0.128) compared to settings with less training data for production ranker. This is consistent with the intuition that stronger rankers determine item positions more accurately based on observable features, allowing $m(\cdot)$ to better predict rankings and leaving less feature-unexplained variation in the residuals. The mean residuals across the 1\%, 10\%, 20\%, and 30\% setups typically fall within a narrow range of difference on each dataset (within roughly 0.02 on Yahoo, 0.01 on MSLR-WEB10k, and 0.015 on Istella-S), indicating that CFC's residual signal varies little across weak-to-moderate production rankers.

Residual variance largely mirrors the mean pattern, shrinking as the production ranker strengthens on Yahoo (0.0088 at 100\%) and MSLR-WEB10k (0.0036 at 100\%). On Istella-S the pattern is less monotonic, but the variance remains below the 20\%--30\% peak. This indicates that stronger rankers generally leave less feature-unexplained variation for the second-stage correction.

All FK statistics are statistically significant at $\bm{p < 0.05}$, supporting the presence of feature-dependent heteroskedasticity consistent with Lewbel’s identification assumption across all production ranker settings and datasets. Higher FK values indicate stronger feature-dependent heteroskedasticity in the residuals. FK magnitudes decrease as the production ranker strengthens, dropping from 81{,}620 to 66{,}574 on Yahoo, from 34{,}178 to 19{,}243 on MSLR-WEB10k, and from 3{,}714 to 2{,}082 on Istella-S as the training fraction increases from 1\% to 100\%. This suggests a reduction in heteroskedasticity under stronger rankers, where residual variance becomes more uniform across the feature space as the production model better captures systematic variation in the feature-dependent data. Despite this reduction, statistically significant heteroskedasticity persists in all settings, ensuring that CFC remains identifiable even under strong production rankers. This is consistent with the robust empirical performance of CFC across production ranker strengths (Table~\ref{tab:prod_variations}).

\begin{table} [hb]
  \fontsize{7.8}{7}\selectfont
  \captionsetup{justification=raggedright, width=0.49\textwidth}
  \caption{A comparison between CFC and CFC-BC using the LambdaMART ranker across various datasets.}
  \vspace{-1em}
  \centering
  \scalebox{1.0}{%
    \begin{tabular}{cccc}
    \toprule
    Datasets & Metrics & CFC & CFC-BC \\
    \midrule

    \multirow{2}{*}{Yahoo}
    & $ERR@10$   & \textbf{0.458} & 0.438 \\
    & $NDCG@10$  & \textbf{0.792} & 0.784 \\
    \midrule

    \multirow{2}{*}{\makecell{MSLR-\\WEB10k}}
    & $ERR@10$   & \textbf{0.321} & 0.315 \\
    & $NDCG@10$  & \textbf{0.490} & 0.481 \\
    \midrule

    \multirow{2}{*}{Istella-S}
    & $ERR@10$   & \textbf{0.715} & 0.699 \\
    & $NDCG@10$  & \textbf{0.762} & 0.754 \\

    \bottomrule
    \end{tabular}
  }
  \label{tab:cfc-with-bc}
\end{table}
\textbf{Impact of using debiased vs. biased clicks for hyperparameter tuning.}
When we assume that true relevance labels are unavailable in the validation dataset, we have click data for hyperparameter tuning, which is inherently biased. For CFC, we use debiased clicks estimated from click data using our proposed click debiasing method (equation~\eqref{eq:valid_clicks}). To evaluate the impact of using debiased clicks versus raw biased clicks for hyperparameter tuning, we define a variant, CFC-BC (CFC with biased clicks), which directly uses the click data for tuning. Comparing CFC and CFC-BC allows us to quantify how debiasing the validation click signal affects ranking performance. Table~\ref{tab:cfc-with-bc} shows the results of CFC and CFC-BC.
CFC achieves the best performance across all datasets, while CFC-BC consistently underperforms. This is because tuning hyperparameters for an unbiased ranking model using biased clicks may not yield the optimal ranker. These results highlight that using biased clicks for validation can lead to suboptimal ranking performance on unbiased test data due to distribution shift.

\balance

\end{document}